\documentclass[prd,letterpaper,tightenlines,nofootinbib,twocolumn,floatfix,showpacs]{revtex4}
\newcommand{\INCLUDEFIGS}{}
\let\originalleft\left
\let\originalright\right
\renewcommand{\left}{\mathopen{}\mathclose\bgroup\originalleft}
\renewcommand{\right}{\aftergroup\egroup\originalright}
\newcommand{\Eq}[1]{Eq.~(\ref{#1})}
\newcommand{\Eqs}[2]{Eqs.~(\ref{#1}) and (\ref{#2})}
\newcommand{\Eqsdash}[2]{Eqs.~(\ref{#1})--(\ref{#2})}
\newcommand{\td}[2]{\frac{d{#1}}{d{#2}}}
\newcommand{\fracd}[2]{\frac{\displaystyle{#1}}{\displaystyle\strut {#2}}}
\newcommand{\recd}[1]{\frac{\displaystyle 1}{\displaystyle\strut {#1}}}
\renewcommand{\r}[1]{(\ref{#1})}
\newcommand{\z}[1]{\left({#1}\right)}

\newcommand{\sz}[1]{\left[{#1}\right]}
\renewcommand{\ae}[1]{\left|{#1}\right|}
\newcommand{\m}[1]{\mathrm{#1}}
\renewcommand{\v}[1]{\mathbf{#1}}
\renewcommand{\c}[1]{\mathcal{#1}}
\newcommand{\rec}[1]{\frac{1}{#1}}
\newcommand{\gvec}[1]{\mbox{\boldmath${#1}$}}
\usepackage{amssymb,graphicx}
\usepackage{amsmath}
\begin{document} 
\setlength{\belowdisplayskip}{8pt plus 2pt minus 5pt}
\setlength{\abovedisplayskip}{8pt plus 2pt minus 5pt}
\title{Simple solutions of fireball hydrodynamics for rotating and expanding triaxial ellipsoids and final state observables}
\author{M.~I.~Nagy$^{1,2}$ and T. Cs{\"o}rg{\H o}$^{3,4}$}
\affiliation{$^1$ELTE, H-1118 Budapest XI, P\'azm\'any P. 1/A,  Hungary}
\affiliation{$^2$Departments of Chemistry and Physics, Stony Brook University, Stony Brook, New York 11794, USA}
\affiliation{$^3$MTA Wigner FK, H-1525 Budapest 114, POB 49, Hungary}
\affiliation{$^4$EKU KRC, H-3200 Gy{\"o}ngy{\"o}s, M\'atrai \'ut 36, Hungary}

\begin{abstract}
We present a class of analytic solutions of non-relativistic fireball hydrodynamics for a fairly general class of equation of state. The presented solution
describes the expansion of a triaxial ellipsoid that rotates around one of its principal axes. We calculate the hadronic final state observables such as
single-particle spectra, directed, elliptic and third flows, as well as HBT correlations and corresponding radius parameters, utilizing simple analytic
formulas. The final tilt angle of the fireball, an important observable quantity, is shown to be not independent of its exact definition: one gets different
tilt angles from the geometrical anisotropies, from the single-particle spectra, and from HBT measurements. Taken together, the tilt angle in the momentum
space and in the relative momentum or HBT variable may be sufficient for the determination of the magnitude of the rotation of the fireball. We argue that
observing this rotation and its dependence on collision energy could characterize the softest point of the equation of state. Thus determining the rotation
may be a powerful tool for the experimental search for the critical point in the phase diagram of strongly interacting matter.
\end{abstract}
\maketitle

\section{Introduction} 

The quest for the experimental investigation of hot and dense strongly interacting matter has always had a fruitful connection to hydrodynamics. The
development of hydrodynamical models that incorporate more and more details about the expansion dynamics of the matter produced in nucleus-nucleus
collisions has been going together hand-in-hand with the richer and richer experimental observations on the particle production mechanism. From the early
days of statistical modelling of multiplicity distributions in high-energy collisions (pioneered by Fermi) through the hydrodynamical description of
rapidity distributions (by the famous Landau-Khalatnikov solution~\cite{Landau:1953gs,Belenkij:1956cd,Khalatnikov} as well as the Hwa-Bjorken
solution~\cite{Hwa:1974gn,Bjorken:1982qr}) nowadays we have various exact analytic as well as numerical solutions of hydrodynamics at hand. These
strive to describe refined observations on essentially three-dimensional momentum spectra (rapidity as well as transverse mass distributions along with
vaious order azimuthal anisotropies), two-particle Bose-Einstein (also named HBT) correlations with resolving power on average momentum, azimuthal angle,
and many other observables. It is impossible to review all these developments here; for a brief summary of hydrodynamical modeling, see e.g.
Ref.~\cite{deSouza:2015ena} and references therein.

A relatively recent research direction in heavy-ion physics phenomenology is to take the rotation of the created matter into account. In non-central heavy-ion
collisions, the non-zero initial angular momentum of the matter influences the time evolution, and numerical modellings of this rotation (of which now there
are many, some pioneering work is to be found in Refs.~\cite{Csernai:2011gg,Csernai:2013vda,Csernai:2013uda}) predict effects on several observables.
According to these models, the effect of rotation can generally be thought of as that of an effective radial flow~\cite{Velle:2015dpa} that influences the
spectra, the elliptic flow as well as two-particle HBT correlations. An equally important prediction was that assuming local thermal equilibrium for spin
degrees of freedom~\cite{Becattini:2013fla}, baryons will be produced with non-zero polarization from a rotating system~\cite{Becattini:2013vja,Xie:2015xpa}.
To observe this, $\Lambda$ baryons are promising candidates, since their polarization can be measured with current experimental setup, by studying their
decay kinematics.

On another notice, it was known for long that particle production from an ellipsoid-like source results in a characteristic oscillation of the HBT radius
parameters as a function of pair azimuthal angle, and if the ellipsoid is tilted in coordinate space by a fixed angle, it results in the appearance of new
cross-terms in the Gaussian approximation of the correlation function. A simple model with these features can be read in Ref.~\cite{Lisa:2000ip}, and a
more advanced, exact hydrodynamical derivation is given in Ref.~\cite{Csorgo:2001xm}. Refs.~\cite{Csernai:2013vda,Csernai:2013uda} also proposed the
differential HBT method to infer the angular momentum of the fireball, although it was pointed out that it is hard to disentangle the effects of rotation
on the HBT radii.

Nowadays one of the most interesting questions in heavy-ion physics concerns the existence (and if it exists, the location) of a critical endpoint on the
phase diagram of strongly interacting matter, as well as the precise experimental determination of the location of the quark-hardon transition on this phase
diagram. Already some finite-size scaling investigations of measured system sizes and freeze-out durations in heavy-ion collisions suggest that the critical
endpoint is in reach with current beam energies at the RHIC accelerator~\cite{Lacey:2014rxa,Lacey:2014wqa}. However, as of now, much additional work is
needed to underpin this statement and to explore the phenomenological properties of this transition. One means to this end is to systematically investigate
the equation of state of the produced matter as a function of beam energy.

The determination of the rotation of the system can give a very useful input to achieve this goal. The importance of rotation --- besides that as an effect
that influences final state observables, it is interesting on its own --- is that the time evolution of the angular velocity of an expanding system depends
on how violent is the expansion, which in turn depends on the equation of state (EoS) of the matter. The main reasons behind this are easy to grasp. On one
hand, if the initial energy density is fixed, then a softer EoS results in less rapid increase of the moment of inertia (because of lower pressure), which
leads to higher angular velocity as compared to the case of a stiffer EoS. Another effect is that adiabatic expansion of a substance with a softer EoS
means slower decrease of the temperature for a given volume change, meaning that there is more time for the system to rotate before reaching the freeze-out
temperature, where the final state observables take their values. In the following we will demonstrate how the interplay of these two effects influence
the final rotation angle of the expanding system.

Rotation is also noteworthy because if one wants to investigate the EoS of the matter using hadronic final state observables, it is important to have
knowledge on the initial conditions of the flow, since different initial conditions and equations of state can lead to similar final states, making the
final state taken alone incapable of determining the EoS. The initial rotational angle of the system can be thought of as either zero or at least as
a monotonic function of the energy of the colliding heavy ions, while the EoS --- and thus the final rotation angle --- is not necessarily monotonic: when
searching for the critical point via the softening of the EoS, it is precisely such a non-monotonic behavior that one is looking for.


The vast work in the field of numerical hydrodynamics and models based on analytic solutions of hydrodynamics supplement each other well. Analytic
models that use exact solutions of both relativistic and non-relativistic hydrodynamics for the description of particle production are naturally harder
to find and more specialized in their initial conditions, but once found, they give general insights in the mechanisms involved in the origin of observables.

As far as we know, the first rotating solutions of relativistic perfect fluid hydrodynamics were found by the simultaneous solution of collisionless
Boltzmann equation~\cite{Nagy:2009eq} and the equations of perfect fluid hydrodynamics. Recent developments concerning exact relativistic and
non-relativistic hydrodynamical solutions with rotation were found also in the framework of AdS/CFT correspondence. Given that high energy heavy ion
collisions and Quark-Gluon Plasmas (QGP-s) produced in these collisions are generically endowed with very large angular momenta, ref.~\cite{McInnes:2014haa} 
proposed to incorporate angular momentum in holographic models. In the case in which the plasma rotates, allowing for a non-vanishing initial angular
momentum usefully improves holographic estimates of the value of the quark chemical potential~\cite{McInnes:2014haa}. Recently it was shown that local
rotation has effects on the QGP at high values of the baryonic chemical potential, which are not only of the same kind as those produced by magnetic
fields, but which can in fact be substantially larger. Furthermore, the combined effect of rotation and magnetism is to change the shape of the
main quark matter phase transition line in an interesting way, reducing the magnitude of its curvature~\cite{McInnes:2016dwk}. Rotation (local vorticity)
and non-vanishing shear stress was also investigated in the solutions found in Refs.~\cite{Hatta:2014gqa} and~\cite{Hatta:2014gga}. These results gave
much thrust to the effort to better understand the rotational expansion of heavy ion collisions and to disentangle the various effects of the rather
high initial angular momentum in these collisions.

In this paper we present a rotating exact solution of the non-relativistic hydrodynamical equations, that is well suited to the geometrical picture of the
strongly interacting matter created in heavy ion collisions. The presented solution features ellipsoidal level surfaces of temperature as well as density,
with three different principal axes. It is a natural generalization of our earlier results presented in Refs.~\cite{Csorgo:2013ksa,Csorgo:2015scx}, where
we explored exact analytic non-relativistic rotating spheroidal solutions (where the two principal axes of the level surface ellipsoids perpendicular to
the rotation is equal to each other), as well as the effect of rotation on the observables. Our new solution also fits into a long line of self-similar but
not rotating solutions of hydrodynamics, both relativistic and non-relativistic ones~\cite{Bondorf:1978kz,Csizmadia:1998ef,Csorgo:2001xm,Csorgo:1998yk,
Csorgo:2001ru,Csorgo:2003ry,Csanad:2012hr,Csanad:2014dpa}. These solutions, as well as the so-called Buda-Lund hydrodynamical
parametrization~\cite{Csorgo:1995bi,Csanad:2003qa}, that gives a reasonable relativistic extension of them, have proven to be adequate tools in the
description of hadronic observables.

The structure of this paper is as follows. In Section~\ref{s:basics} we recite the hydrodynamical equations suited for the treatment of our problem at hand.
It turns out that the solution that we are after can be easily written up in a rotating reference frame instead of the laboratory frame of the colliding
nuclei. In Section~\ref{s:solution} we present the solution for an expandig rotating triaxial ellipsoid (i.e. an ellipsoid with three different principal
axes): Section~\ref{ss:hydrosol} presents the solution itself in a compact way (those interested only in this should look up this section), while discussion
is left to Sections~\ref{ss:analysis} and \ref{ss:conserved}, and technical details about the derivation is left to Appendix~\ref{s:app:generalsol}. In
Section~\ref{s:obs} we calculate the final state hadronic observables (spectra, flow parameters, HBT correlation function) using simple analytical formulas.
These formulas enable us to draw some general conclusions on the effect of rotation on the observables. In Section~\ref{s:discussion} we illustrate the
time evolution of the system as well as the effect of rotation on the observables using some simple and reasonable initial conditions. The detailed
investigation of the available experimental data is beyond the scope of this paper, however, we point out that the simultaneous measurement of the harmonic
flow parameters ($v_1$, $v_2$, $v_3$) and the azimuthal oscillation of HBT radii (especially the cross-terms in the so-called Bertsch-Pratt parametrization)
gives a means to determine the angular velocity as well as the final tilt angle of the ellipsoidal expanding system, providing a path to determine the
softest point of the equation of state as outlined above. Finally we summarize and conclude.

\section{Basic equations}\label{s:basics}
\subsection{Equations of hydrodynamics}\label{ss:hydroeqs}

We outline the non-relativistic hydrodynamical equations in a form suited to our task of finding rotating exact solutions. The fluid motion is described by
the velocity field $\v v$, the pressure $p$, the energy density $\varepsilon$, the temperature $T$, the particle number density $n$, the chemical potential
$\mu$, and the entropy density $\sigma$. All these hydrodynamical quantities are functions of $t$ and $\v r$, the time and the spatial coordinate. The
fundamental equations are the particle number and energy conservation equations as well as the Euler equation:
\begin{align}
\partial_tn + \nabla\z{n\v v}   &= 0 , \label{e:cont} \\
\partial_t\varepsilon + \nabla\z{\varepsilon\v v} &= - p\z{\nabla\v v} . \label{e:energy} \\
\partial_t\v v + \z{\v v\nabla}\v v &= -\nabla p / \z{m_0 n} . \label{e:Euler}
\end{align}
Here $m_0$ is a the mass of an individual particle. Using the well-known thermodynamical relations
\begin{align}
\varepsilon+p &= T\sigma + \mu n,\\ 
\m d\varepsilon &= T\m d\sigma + \mu\m dn,
\end{align}
one can verify that the energy conservation equation \Eq{e:energy} is equivalent to the entropy conservation:
\begin{equation}\label{e:entropy}
\partial_t\sigma + \nabla\z{\sigma\v v} = 0.
\end{equation}
This set of equations need to be supplemented by an appropriate equation of state providing
a relation between $T$, $p$ and $\varepsilon$. Just as in Refs.~\cite{Csorgo:2001xm,Csorgo:2013ksa}, we choose
\begin{align} 
p           &= n T,\\
\varepsilon &= \kappa(T) p .\label{e:eos} 
\end{align}
This EoS is thermodynamically consistent for any $\kappa(T)$ function, as was shown e.g. in
Ref.~\cite{Csorgo:2001xm}. It is a generalization of the case for constant $\kappa$, which would correspond to a non-relativistic ideal gas for
$\kappa = 3/2$, and to an ultra-relativistic ideal gas for $\kappa = 3$. The arbitrary $\kappa(T)$ function introduced here allows one to incorporate
any temperature dependent speed of sound $c_s^2 = dp/d\epsilon = 1/\kappa(T)$.

Just as in Ref.~\cite{Csorgo:2013ksa}, we may rewrite \Eqsdash{e:cont}{e:Euler} for the independent variables $T$, $n$ and $\v v$ as follows:
\begin{align}
\z{\partial_t+\v v\nabla}n &= -n\nabla\v v, \label{e:ncont2} \\
\sz{T\td{\kappa}{T}+\kappa}\z{\partial_t+\v v\nabla}T &= -T\nabla\v v, \label{e:Tcont2} \\
nm_0\z{\partial_t+\v v\nabla}\v v &= - n\nabla T - T\nabla n . \label{e:Euler2}
\end{align}
Also, following Refs.~\cite{Csorgo:2013ksa,Csorgo:2015scx}, we note here that this set of equations is valid for the case when there is non-vanishing
$n$ particle density which embodies the fact that there is a meaningful total particle number that is conserved. This assumption is valid for the late
stages of the hydrodynamic evolution of the matter produced in heavy-ion collisions, when the kinetic freeze-out is not yet reached but the particle
type changing hadronic reactions ceased to play a role. For the case generally thought to apply to the quark-gluon-plasma phase, that is, when there
is no conserved particle density, we may (again following Ref.~\cite{Csorgo:2013ksa}) write up a separate set of hydrodynamic equations, the main
difference being that here the only independent variables are $T$, $\sigma$ and $\v v$, and the mass term in the Euler equation is different: 
\begin{align}
\partial_t\sigma + \nabla\z{\sigma\v v} &= 0 , \label{e:caseBentropy} \\
\z{\varepsilon+p}\z{\partial_t\v v + \z{\v v\nabla}\v v} &= -\nabla p , \label{e:caseBeuler}
\end{align}
which, using the thermodynamical relations $\varepsilon+p = T\sigma$, $\m dp = \sigma\m dT$ (which are valid for $n = 0$) are rewritten as
\begin{align}
\z{\partial_t+\v v\nabla}\sigma &= -\sigma\nabla\v v ,\label{e:caseBentropy2} \\
T\z{\partial_t+\v v\nabla}\v v &= -\nabla T . \label{e:caseBeuler2}
\end{align}
The mass term $\varepsilon+p$ in the Euler equation (the enthalpy density) stems from the relativistic version of the Euler equation. In the non-relativistic
case with a conserved particle number, one is led to make the approximation $\mu \approx m_0$, and thus $\varepsilon + p = T\sigma + \mu n \approx m_0n$.
The case for vanishing $n$ is the opposite limiting case, when the mass term stems entirely from the entropy density.

The basic equations for vanishing $n$, \Eqs{e:caseBentropy2}{e:caseBeuler2} also have to be supplemented with an EoS. The convenient
choice again is simply
\begin{equation}\label{e:EoScaseB}
\varepsilon = \kappa(T)p \quad\Leftrightarrow\quad \sz{\kappa(T)+1}p = T\sigma .
\end{equation}
With an appropriate $\kappa(T)$ function, one can describe e.g. the equation of state of the strongly interacting matter inferred from lattice QCD
calculations.

As seen already in Ref.~\cite{Csorgo:2013ksa}, the solution of these two sets of equations (one valid for non-vanishing $n$, the other for vanishing $n$)
can be done very similarly to each other; this is also true for the solutions presented in this paper. In the following, we mainly restrict ourselves
to the case when there is a conserved $n$, i.e. to the solution of \Eqsdash{e:cont}{e:Euler}, mainly because we want to calculate the final state hadronic
observables which are formed in the final states of the hydrodynamical evolution, where this approximation is thought to be valid.

\subsection{Equations in a rotating reference frame}\label{ss:rotatingframe}
For our treatment, the shape of the hot and dense matter that is created in non-central heavy-ion collisions can be approximated with a triaxial
ellipsoid that has non-zero angular momentum, and also expands violently. As customary in heavy-ion phenomenology, let the $z$ axis point in the
direction of the incoming projectiles, and the $x$ axis point in the direction of the impact parameter. In the following this inertial frame is
called the laboratory frame, denoted by $K$. The rotation is assumed to be in the $x$--$z$ plane, around the $y$ axis.

It turns out that finding a solution which describes the physical situation of interest to us, i.e. a triaxial, expanding and simultaneously rotating
ellispoid, is simpler to achieve in a frame which rotates together with the expanding ellipsoid. This frame is denoted by $K'$, with its axes,
$x'$, $y'$ and $z'$, pointing in the directions of the principal axes. The $y'$ axis is the same as the $y$ axis. We denote the rotation angle of $K'$ with
respect to $K$ in the $x$--$z$ plane by $\vartheta(t)$. (We will sometimes omit the explicit notation of the time dependence for functions introduced as
functions of $t$.) We introduce the rotation matrix $\v M$ that connects the $K$ and $K'$ frames, and also the vector $\gvec\Omega$ as the angular velocity
of $K'$ with respect to $K$:
\begin{equation}\label{e:Mv}
\v M(t) \equiv \begin{pmatrix} \cos\vartheta & 0 & -\sin\vartheta \\ 0 & 1 & 0 \\ \sin\vartheta & 0 & \cos\vartheta \end{pmatrix},\quad
\gvec\Omega = \begin{pmatrix} 0 \\ \dot\vartheta \\ 0 \end{pmatrix} ,
\end{equation}
so the coordinate and the velocity components transform between $K$ and $K'$ as
\begin{equation}\label{e:vcommaOmega}
\v r' = \v M(t)\v r,\quad 
\v v' = \v M\v v-\gvec\Omega\times\v r' .
\end{equation}

Of \Eqsdash{e:ncont2}{e:Euler2} or \Eqs{e:caseBentropy2}{e:caseBeuler2}, the continuity-like equations retain their form in $K'$, but the Euler equation
needs to be supplemented with inertial force terms. The basic equations in the $K'$ frame are then
\begin{align}
\z{\partial'_t+\v v'\nabla'}n &=-n\nabla'\v v',\label{e:rotncont}\\
\sz{T\td{\kappa}{T}+\kappa}\z{\partial'_t+\v v'\nabla'}T &= -T\nabla'\v v', \label{e:rotTcont}\\
\z{\partial'_t+\v v'\nabla'}\v v' &= -\frac{\nabla' T}{m_0} - \frac{T}{n}\frac{\nabla' n}{m_0} + \v f' ,\label{e:roteuler}
\end{align}
\begin{equation}\label{e:inertialfdef}
\v f'\equiv2\v v'\times\gvec\Omega+\gvec\Omega\times\z{\v r'\times\gvec\Omega}+\v r'\times\dot{\gvec\Omega}.
\end{equation}
The terms in $\v f'$ describe the Coriolis force, the centrifugal force and the force stemming from the angular acceleration
of the $K'$ frame. We introduced the $\nabla'$ and $\partial'_t$ notations for derivatives in the $K'$ frame: $\nabla'$ means derivatives
with respect to the $\v r'$ coordinates, while $\partial'_t$ is time derivative for $\v r'$ fixed. (This is different from $\partial_t$,
because the relation between $\v r'$ and $\v r$ is time-dependent.)

\section{Rotating ellipsoidal solutions}\label{s:solution}
The solution presented below is a direct generalization of earlier results describing non-rotating ellipsoidal expansion~\cite{Csorgo:2001xm}, as well as
rotating solutions~\cite{Csorgo:2013ksa}: many features carry over essentially unchanged into our treatment, and we conform our notations to those used
in these works. In the following section, we present the solutions in a concise form; additional technical details can be found in
Appendix~\ref{s:app:generalsol}. In order to enhance the clarity and transparency of the presentation, we also provide Appendix~\ref{s:app:labsol},
where we summarize our new exact solutions in the laboratory (inertial) frame. 

\subsection{New rotating triaxial solutions}\label{ss:hydrosol}

As mentioned, the new solutions are easier to write up in the co-rotating $K'$ frame. From
\Eqs{e:Mv}{e:vcommaOmega}, the relations between the coordinate and the velocity components in $K$ and $K'$ are
\begin{align}
r'_x &= r_x\cos\vartheta-r_z\sin\vartheta,\label{e:rxtransform}\\
r'_z &= r_x\sin\vartheta+r_z\cos\vartheta,\\
v'_x &= v_x\cos\vartheta-v_z\sin\vartheta-\dot\vartheta r'_z,\label{e:vxtransform}\\
v'_z &= v_x\sin\vartheta+v_z\cos\vartheta+\dot\vartheta r'_x.\label{e:vztransform}
\end{align}
The $y$ components do not mix: $v'_y = v_y$, $r'_y = r_y$.

Following the mentioned earlier works, we introduce the time-dependent principal axes of the rotating ellipsoid, $X(t)$, $Y(t)$, $Z(t)$. We also introduce
the \emph{scaling variable} $s$, whose level surfaces correspond to the rotating ellipsoidal level surfaces of the temperature and density, and the
characteristic volume $V$ and average lateral radius $R$ of these ellipsoids:
\begin{equation}\label{e:sdef}
s = \frac{{r_x'}^2}{X^2} +\frac{{r_y'}^2}{Y^2} +\frac{{r_z'}^2}{Z^2} ,\quad V \equiv (2 \pi)^{3/2} XYZ ,\quad R\equiv \frac{X+Z}{2}.
\end{equation}
In order to obtain the desired rotating solution, we specify the $\dot\vartheta(t)$ quantity and the velocity field by introducing
the $\omega(t)$ ``angular velocity'' as follows:
\begin{equation}
\dot\vartheta(t)\equiv \frac{\omega(t)}{2},\quad \omega(t)=\omega_0\frac{R_0^2}{R^2(t)},\label{e:omegadef}
\end{equation}
\begin{equation}\label{e:vdef}
\v v'\z{\v r',t} = \z{\begin{array}{l}
{\frac{\dot X(t)}{X(t)}r'_x + \frac{\omega(t)}{2}\frac{X(t)}{Z(t)}r'_z}\vspace{2mm} \\
{\frac{\dot Y(t)}{Y(t)}r'_y}\vspace{2mm} \\
{\frac{\dot Z(t)}{Z(t)}r'_z - \frac{\omega(t)}{2}\frac{Z(t)}{X(t)}r'_x} \end{array}}.
\end{equation}
This velocity field preserves the $s=$const ellipsoids, as $s$ is constant along the trajectories of the fluid elements:
\begin{equation}\label{e:tdst}
\z{\partial_t+\v v\nabla}s = 0\quad\Leftrightarrow\quad \z{\partial'_t+\v v'\nabla'}s = 0 .
\end{equation}

The \r{e:rotncont}--\r{e:rotTcont} density and temperature equations are solved along similar lines as in e.g.
Ref.~\cite{Csorgo:2013ksa}. We distinguish two cases:
\begin{itemize}
\item\emph{Case A:} If we assume for the \r{e:eos} EoS that $\kappa(T)=\kappa=$const, we can have the solutions
\begin{equation}\label{e:nTsolA}
n\z{\v r',t} = n_0\frac{V_0}{V}\nu(s),\quad T\z{\v r',t} = T_0\z{\frac{V_0}{V}}^{\rec\kappa}\c T(s)
\end{equation}
for \Eqs{e:rotncont}{e:rotTcont}.
Here $V_0$ is the initial value of the volume $V$, and (just as in e.g. Refs.~\cite{Csorgo:1998yk,Csorgo:2001ru}) the $\nu(s)$ and $\c T(s)$
functions obey the condition
\begin{equation}\label{e:nutau}
\nu(s) = \rec{\c T(s)}\exp\z{-\rec 2\int_0^s\frac{\m ds'}{\c T(s)}},
\end{equation}
so only one of them can be chosen independently. (In e.g. Ref.~\cite{Csorgo:2001ru} the similar condition is expressed with an additional free
scale parameter introduced; it can be absorbed into the scales of the $X$, $Y$, $Z$ axes.) \Eq{e:nTsolA} describes an adiabatic expansion, where
the familiar $T^\kappa V=$const relation holds. The coordinate dependence of $n$ and $T$ enters only through $s$, so these profiles are self-similar.

\item
\emph{Case B:} If we allow any temperature dependent $\kappa(T)$ function in the \r{e:eos} equation of state, then the relevant solution for
\Eqs{e:rotncont}{e:rotTcont} is specified by a Gaussian density profile and a spatially homogeneous temperature profile:
\begin{equation}\label{e:nTsolB}
n\z{\v r',t} = n_0\frac{V_0}{V}e^{-s/2},\quad T\z{\v r',t} \equiv T(t).
\end{equation}
The time evolution of $T(t)$ is given by the following differential equation stemming from \Eq{e:rotTcont}:
\begin{equation}
\frac{\m d \sz{ T \kappa(T)} }{\m d T} \frac{\dot T}{T} + \frac{\dot V}{V} = 0
\end{equation}
which can be integrated in an implicit relation that yields the volume as a function of the temperature T
\begin{equation}\label{e:Tdiffe}
\ln\frac{V_0}{V}=\int\limits_{T_0}^T\frac{\m dT'}{T'}\td{\sz{T'\kappa(T')}}{T'} .
\end{equation}
\end{itemize}
Cases A and B have a common special case if $\kappa(T)=\kappa=$const and $\nu(s)=e^{-s/2}$: if $\kappa(T)=$const, then \Eq{e:Tdiffe} can be solved for
$T(t)$ to yield the form in \Eq{e:nTsolA}, and if $\c T(s)=1$, then \Eq{e:nutau} implies a Gaussian density profile:
\begin{equation}\label{e:gaussn}
\c T(s)=1\quad\Rightarrow\quad \nu(s)=e^{-s/2}.
\end{equation}
To get a full solution of the Euler equation, the time evolution of the principal axes $X$, $Y$, $Z$ must obey a set of ordinary differential
equations. For both Case A and B, they can be written up in the short form
\begin{equation}\label{e:eqmotXYZkappa}
X\big(\ddot X-\omega^2R\big) = Y\ddot Y = Z\big(\ddot Z -\omega^2R\big) = \frac{T}{m_0}.
\end{equation}
The formulas in this subsection describe a rotating, triaxial, expanding fireball and they correspond to an exact solution of hydrodynamics: it can be
directly verified that they indeed solve \Eqsdash{e:rotncont}{e:roteuler}.

\subsection{Analysis of the new solutions}\label{ss:analysis}

The meaning of the equations of motion in \Eq{e:eqmotXYZkappa} is that the hydrodynamical problem is reduced to a set of ordinary differential equations.
Although a general analytical solution to these ordinary differential equations is lacking, in terms of the hydrodynamical problem, they can be considered
as readily solvable for any initial conditions, at least numerically. In this sense, our new
solutions can be called \emph{parametric} ones, just as those found in Refs.~\cite{Csorgo:2001ru,Csorgo:2001xm,Csorgo:2013ksa,Csorgo:2015scx}. The equations
of motion encountered here are also natural generalizations of those found in these earlier works. It must be remembered, however, that our new equations
are valid for the axes in the rotating $K'$ frame. In our new class of solutions, there are eight independent initial conditions: the initial values
$X_0$, $Y_0$, and $Z_0$ of the principal axes, their initial time derivatives $\dot X_0$, $\dot Y_0$, and $\dot Z_0$, as well as $T_0$, the initial
temperature in the centre of the fireball, and the $\omega_0$ parameter that quantifies the initial value of the angular velocity, characterizing the
rotation around the $y$ axis.

We introduced the ``average'' angular velocity $\omega(t)$ of the flow, as well as the ``average radius'' $R(t)$ (and its
initial value $R_0$) to conform with the earlier spheroidal solutions of Ref.~\cite{Csorgo:2013ksa}; this notation will be useful in the following.
In the spheroidal limiting case $X=Y=R$ will hold, and this $R$ is the same as the radial size of the ellipsoid in the case of the spheroidal solution.

Also we note that of the so-called vorticity of the flow, $\gvec\omega\z{\v r,t}\equiv \nabla\times\v v\z{\v r,t}$, only the $y$-component is non-vanishing,
and it takes the simple form of
\begin{equation}\label{e:vorticity}
\gvec\omega_y\z{\v r,t}\equiv \z{\nabla\times\v v\z{\v r,t}}_y = \omega(t)\frac{(X+Z)^2}{2XZ} .
\end{equation}
In the case of $X=Z$, $\gvec\omega_y = 2\omega(t)$, just as in Ref.~\cite{Csorgo:2015scx}. We will also see that this $\omega(t)$ is the quantity
characterisitic to the rotation that appears in the expression of the observables in a straightforward way.

To elucidate the characteristic of our velocity field, we note again that taken in the inertial $K$ frame by substituting \Eq{e:vdef} into
\Eqs{e:vxtransform}{e:vztransform}, it reduces to the velocity field of Refs.~\cite{Csorgo:2013ksa} in the $X=Y=R$ case. But it must be noted here
that in the case of our new general, $X\neq Z$ solution, it is \emph{not} simply the case that we have a rotating frame $K'$ which is the eigenframe of
the rotating ellipsoidal surfaces, and also the velocity field is non-rotating in the $K'$ frame. We see from \Eq{e:vdef} that the expression of $\v v'$
in the $K'$ frame also involves rotational, off-diagonal terms: this means that these off-diagonal terms and the rotation of $K'$ with respect to $K$
each ``carry'' half of the rotation of the velocity field. In other words, the rotation of the velocity field (in
the inertial $K$ frame) has two ``components'': first, the expanding ellipsoidal surfaces (to which the $K'$ frame is fixed) are rotating in the
$K$ frame, governed by $\dot\vartheta(t)$ as seen from \Eqs{e:vxtransform}{e:vztransform}, and secondly, the velocity field rotates with respect to the
$K'$ frame, as seen from \Eq{e:vdef}. It turns out after some investigation of the Euler equation that one cannot find a solution where the
$\v v'$ velocity field is diagonal (i.e. non-rotating) in the frame fixed to the rotating ellipsoids (i.e. the $K'$ frame). In
Appendix~\ref{s:app:generalsol} we get back to this question.

For the principal axes $X$, $Y$, $Z$, the equations of motion were written up in \Eq{e:eqmotXYZkappa}. This form suits both cases, Case A and B as discussed
in the previous subsection: in Case B, for arbitrary $\kappa(T)$ function but spatially homogeneous $T$ profile, the r.h.s. of \Eq{e:eqmotXYZkappa} is
just the $T(t)$ function, which is in turn determined as a function of $X(t)$, $Y(t)$, and $Z(t)$ (through the volume $V=XYZ$) implicitly by
\Eq{e:Tdiffe}. In case of $\kappa(T) = \kappa = $const, $T(t)$ has an explicit form as given by \Eq{e:nTsolA}. As discussed already, the constant
$\kappa$ case allows a more general, coordinate-dependent temperature, given by \Eq{e:nTsolA}. The equations of motion of the axes in this case
are again \Eq{e:eqmotXYZkappa}, in the sense that the $T$ on the r.h.s. is to be understood as the ``time-dependent part'' of $T(t,\v r')$, i.e.
if one wrote $\c T(s)=1$ in \Eq{e:nTsolA}. 

The equations of motion for the principal axes $X$, $Y$, $Z$ can be thought of as the equations of motion of a particle with mass $m_0$ in an external
potential. We write up the Hamiltonian governing this motion corresponding to \Eq{e:eqmotXYZkappa} only in the constant $\kappa$ case (Case A in
Section~\ref{ss:hydrosol}): 
\begin{eqnarray}
\label{e:Hamilton}
H &=& \rec{2m_0}\z{P_X^2+P_Y^2+P_Z^2} +U\z{X,Y,Z}, \\
U\z{X,Y,Z} & = &
\kappa T_0\z{\frac{X_0Y_0Z_0}{XYZ}}^{1/\kappa} + \frac{ m_0\omega_0^2}{4}\frac{(X_0+Z_0)^4}{(X+Z)^2} . \nonumber \\
\null & \null & \null
\label{e:HamiltonU}
\end{eqnarray}
The momenta $P_X$, $P_Y$, $P_Z$ are just equal to $m_0\dot X$, $m_0\dot Y$, $m_0\dot Z$, respectively. The potential term can be also written in a
short-hand notation, with the help of $R$ and $\omega$ as defined in \Eq{e:omegadef}, as
\begin{equation}\label{e:HamiltonU-short}
U = \kappa T + m_0\omega^2R^2 .
\end{equation}
In the case of a non-constant $\kappa(T)$, the Hamiltonian that gives back \Eq{e:eqmotXYZkappa} also can be written up in a much similar way,
the only difference is the form of the temperature related term in the expression of the potential $U$. We do not indulge in this now, but mention that
this can be done in a way similar to that outlined in Ref.~\cite{Csorgo:2013ksa}.

If we set $X=Z\equiv R$ in our equations, we get back the results of Ref.~\cite{Csorgo:2013ksa}: we see that in this case indeed the $\omega_0$ plays the
role of the initial value of the angular velocity of the fluid. In the case of $X=Z$, the meaning of the $\vartheta$ angle becomes ill-defined: for
a rotating spheroid one clearly cannot uniquely define the tilt of the co-rotating coordinate system \emph{and} the rotational velocity with respect
to that frame separately. In the spheroidal case only the total angular velocity of the fluid (i.e. that with respect to the inertial $K$ frame) has
a definite meaning; as discussed above, in some sense half of this angular velocity is provided by the rotation of $K'$, the other half by the
rotation of $\v v'$ in $K'$. 

\subsection{Conserved quantities}\label{ss:conserved}

It is also worthwhile to calculate some conserved quantities. We do this only for the case of constant $\kappa$, and for simplicity we also specify the
spatial shape of the density $n$ and $T$ by taking the spatially homogeneous temperature and Gaussian density case, \Eq{e:gaussn}. In this case, the
total particle number $N_0$ is
\begin{equation}\label{e:N0}
N_0 = \int\m d^3\v r' n\z{t,\v r'} = 
              n_0V_0,
\end{equation}
which is clearly a constant.

The total (kinetic and internal) energy $E_0$ of the fluid turns out to be
\begin{equation}\label{e:E0}
E_0 = \frac{m_0}{2}\z{\dot X^2+\dot Y^2+\dot Z^2} + U\z{X,Y,Z},
\end{equation}
with $U\z{X,Y,Z}$ given by \Eq{e:HamiltonU}. This is precisely the value of the Hamiltonian; the conservation of $E_0$ is thus equivalent to the
Hamiltonian formulation of the equations of motion for $X$, $Y$, $Z$.

In the special case of constant $\kappa(T) = \kappa = 3/2$ (which is the case of a non-relativistic ideal gas),
one can write up another first integral of \Eq{e:eqmotXYZkappa}. Combining these equations with the energy conservation
equation obtained from the $H =$const criterion, with $H$ defined in \Eq{e:Hamilton}, for $\kappa = 3/2$ one gets the following solution for the time
evolution of $X^2+Y^2+Z^2$:
\begin{eqnarray}
\null &\null &
X^2+Y^2+Z^2 = \frac{2E_0t^2}{m_0N_0} + 2\sz{X_0\dot X_0+Y_0\dot Y_0+Z_0\dot Z_0}t + \nonumber \\
\null & \null & +X_0^2+Y_0^2+Z_0^2,\quad\m{if}\quad\kappa = 3/2,
\label{e:X2Y2Z2}
\end{eqnarray}
with the initial conditions taken at $t=0$. This is similar to the results found in e.g. Refs.~\cite{Csorgo:2015scx,Akkelin:2000ex}.

Another important quantity is the total angular momentum $\v J_0$\ of the fluid. In our setting, only the $J_y$ component is non-zero: its value turns
out to be
\begin{equation}\label{e:Jy}
J_y = 2N_0m_0\omega_0R_0^2,\quad R_0\equiv \frac{X_0+Z_0}{2},
\end{equation}
which is indeed constant. To frame this expression, we can calculate the (time-dependent) moment of inertia $\Theta(t)$ of the fluid (with respect
to the $y$ axis). For the Gaussian shape of $n$ specified by \Eq{e:gaussn}, we get 
\begin{align}
\Theta(t) & = m_0N_0\z{X^2+Z^2}, \label{e:Theta}\\
J_y & = \Theta(t)\omega(t)\times\frac{2R^2}{X^2+Z^2}. \label{e:JTheta}
\end{align}
We see that the analogy to the spherical case is not complete; our new solution has a richer structure. Also, the rotational motion in our solution is
clearly different from that of a rotating solid body. Nevertheless, again in the $X=Z\equiv R$ special case, the present formulas give back those valid
for the spheroidal case.
\begin{center}*$\quad$*$\quad$*\end{center}
In some sense the solution presented above is a fairly general self-similar rotating solution. In Appendix~\ref{s:app:generalsol} we outline the reasoning
that leads to this solution, and show that from the ansatz specified up to now, a slightly more general solution also may follow. However, for the problem
under consideration, namely, the rotating expansion of the fireball produced in heavy-ion collisions, the additional generality in that solution is
apparently irrelevant.

It should be also emphasized that (as seen from the treatment of the problem in Appendix~\ref{s:app:generalsol}) although one would very much prefer a
rotating solution which is a more direct generalization of the previously known ellipsoidal or rotating spherical solutions, i.e. where the ellipsoids
co-rotate with the velocity field (in the sense that $\v v'$ is a non-rotating diagonal one in $K'$, so the cross-terms in \Eq{e:vdef} are missing),
such solutions simply do not exist. In this sense the solution presented here is the simplest one corresponding to triaxial rotating ellipsoids. 

\section{Calculation of hadronic observables}\label{s:obs} 

Having seen a hydrodynamical solution whose time dependence mirrors that of an expanding rotating triaxial ellipsoid, that is a reasonable analytic
model for the rotating expanding time evolution of the strongly interacting matter created in heavy-ion collisions, we now turn to the question of
what observable quantities carry information on the rotation of the system. It turns out that the hadronic observables for the considered solution
can be expressed by means of simple analytic formulas. In this section we outline these calculations
and discuss what observables are sensitive to the rotation.

In the usual way in hydrodynamical modelling, we assume that the system (the fluid) freezes out on some hypersurface, i.e. the hydrodynamical evolution
abruptly stops, to give way to the final observable particles. The phase-space distribution of the system at the instant of freeze-out then determines
the final state distributions. We specify the solution that we will investigate as well as the freeze-out condition in the simplest way that suits
the calculation: we take the spatially homogeneous temperature case (with a Gaussian density profile, \Eq{e:gaussn}), and assume that the freeze-out
sets in at a given $T_f$ temperature. In our $T\equiv T(t)$ case this also means that freeze-out is happening at a given time, $t_f$, everywhere
simultaneously. We assume that at the freeze-out, particles with mass $m$ appear. Throughout the calculation, $m$ is retained as a free parameter,
however, in practical cases, the mass of the produced particles may be taken the fixed values valid for e.g. pions, kaons, or (anti)protons
($m = 140$ MeV, $m=494$ MeV, and $m=938$ MeV, respectively). An estimation of $T_f$ was made already by Landau and Belenkij~\cite{Belenkij:1956cd}:
$T_f\approx m_\pi$, the pion mass, since this is the typical energy at which the hadronic collisions that transform particle types cease,
and thus this is the typical temperature at which the mean free path of a pion gas starts to increase exponentially.

It may be also noted that our treatment is fully non-relativistic, an assumption that
allows a fully analytic calculation, but questionable as a realistic assumption for intermediate transverse momenta. Concerning relativistic
parametrizations, an easily performed generalization of the exact formulas stemming from non-relativistic solutions is described in the framework
of the so-called Buda-Lund model~\cite{Csorgo:1995bi,Csanad:2003qa}. The end result here is basically that one may substitute $m_t = \sqrt{p_t^2+m^2}$
(transverse mass) into the place of the mass $m$ of the individual particles, for a rudimentary relativistic generalization. So the $m$ dependence
in our following formulas for the observables may be understood as a preliminary suggestion on the $m_t$ dependence what one might get in a more
realistic relativistic treatment. However, in this paper we only deal with fully analytic (thus in our case, non-relativistic) formulas, with
$m$ being the mass of the particle.

\subsection{Source function}\label{ss:sourcefunction}
The observables are calculated from the \emph{source function} or \emph{emission function}, denoted by $S\z{t,\v r,\v p}$, the thermal phase-space
distribution taken at the freeze-out time, $t_f$. It can be written in our non-relativistic approximation, for our case of solution, as
\begin{equation}\label{e:Sdef}
S\z{\v{r},\v{p}}\propto \frac{n\z{t_f,\v r}}{T_f^{3/2}}\exp\left\{-
\frac{\z{\v p -m\v v\z{t_f,\v r}}^2}{2mT_f}\right\},
\end{equation}
with every hydrodynamical quantity taken at the freeze-out time. This is normalized so that the integral over $\v p$ at a given point
$\v{r}$ is proportional to the number density, $n$ at that point. Here $m$ stands for the mass of the produced particle which may or may not be
equal to the $m_0$ parameter that governs the time evolution of the hydrodynamical evolution as in \Eq{e:eqmotXYZkappa}.

One can use this source function to calculate two important set of observables: the single particle spectrum (and its
corollaries, like azimuthal anisotropies), and two-particle correlation functions (and related quantities, like HBT radii).
The defining formula of the single-particle spectrum $N_1\z{\v p}$ in our hydrodynamical setting is
\begin{equation}\label{e:N1pdef}
N_1\z{\v p} \equiv E\frac{dn}{d^3\v{p}} \propto E\int\m d^3\v r\,S\z{\v p,\v r},
\end{equation}
where $E$ is the particle energy.

Bose-Einstein or HBT-correlations of bosons stem from their quantum mechanical indistinguishability, and in turn, the symmetry property of
their wave-function. Assuming interaction-free final state, the two-particle Bose-Einstein correlation function $C\z{\v K,\v q}$ is
connected to the Fourier transform of the emission function:
\begin{equation}\label{e:C2def}
C\z{\v K,\v q} \approx 1 + \lambda\frac{\big|\tilde S_{\v K}\z{\v q}\big|^2}{\big|\tilde S_{\v K}\z{\v 0}\big|^2},
\end{equation}
with $\v K = \rec 2\z{\v p_1+\v p_2}$ being the average momenutm and $\v q = \v p_1-\v p_2$ the relative momentum of the pair, and
\begin{equation}\label{e:tildeS}
\tilde S_{\v K}\z{\v q} = \int\m d^3\v r\,e^{i\v q\v r}S\z{\v r,\v K} ,
\end{equation}
with $S\z{\v r,\v p}$ taken at the average momentum $\v K$. The so-called intercept parameter $\lambda$ measures the correlation strength at zero momentum.
A phenomenological explanation for $\lambda$ is the so-called core-halo model~\cite{Csorgo:1994in}, where $\sqrt\lambda$ measures the ratio of
primordial particles (pions) to all the produced ones (including those coming from long-lived resonance decays). The approximation in \Eq{e:C2def} is,
among other things, that one writes $\v K$ in the argument of the source function in \Eq{e:tildeS}, and also that one neglects multi-particle correlation
effects, correlated particle production, and Coulomb interactions (this latter one can be straightforwardly corrected for).

In what follows, we outline the calculations that lead to our results on the observables. The calculations are in essence very simple, since
they involve only Gaussian integration, albeit multivariate Gaussians with mixed second-order terms.
Using \Eqs{e:gaussn}{e:vdef}, we clearly see that indeed $S\z{\v r,\v p}$ is Gaussian in the coordinates.

Our hydrodynamical solution outlined in Section~\ref{s:solution} was written up in a rotating reference frame, $K'$, whose tilt angle with respect to the
inertial $K$ frame, $\vartheta(t)$ was one of the dynamical variables of the rotating expansion. We got an expression for $\dot\vartheta(t)$ that can 
be numerically integrated to yield the final tilt angle $\vartheta_f\equiv \vartheta\z{t_f}$. The calculation of the observables is most easily
done in an \emph{inertial}, i.e. non-rotating reference frame that is tilted with $\vartheta_f$ with respect to $K$. We may denote this frame by
$\overline K'$: the momentum components in this frame are related to the $K$-components similarly as the coordinate $\v r'$ to $\v r$:
\begin{equation}\label{e:ppcomma}
\overline{\v p'} = \v M\v p \quad\Leftrightarrow\quad
\begin{array}{rl}
p'_x &= p_x\cos\vartheta-p_z\sin\vartheta,\\
p'_z &= p_x\sin\vartheta+p_z\cos\vartheta,
\end{array}
\end{equation}
with $\v M$ defined in \Eq{e:Mv}. However, the velocity field $\overline{\v v}'$ in the $\overline K'$ frame is different from $\v v'$ as
introduced in \Eq{e:vcommaOmega} precisely because the $\overline K'$ frame is inertial, so $\overline{\v v}'$ does \emph{not} contain the effect
of angular motion:
\begin{equation}\label{e:overlinev}
\overline{\v v'} = \v M\v v \quad\Leftrightarrow\quad
\begin{array}{rl}
\overline v'_x &= v'_x+\dot\vartheta r'_z,\\
\overline v'_z &= v'_z-\dot\vartheta r'_x.
\end{array}
\end{equation}
The coordinates in the $\overline K'$ frame are of course the same as those in $K'$, with the components of that of $\v r'$.
Again, the $y$ components do not mix: $p'_y = p_y$, $\overline v'_y = v'_y = v_y$.

We need to plug these expressions into \Eq{e:Sdef}. So the final expression of the source function is
\[
S\z{\v r',\v p'}\propto \frac{n_0}{T_f^3}\exp\z{-\frac{{r'}_x^2}{2X_f^2}-\frac{{r'}_y^2}{2Y_f^2}-\frac{{r'}_z^2}{2Z_f^2}}\times
\]
\begin{equation}\label{e:Srexpr}
\times\exp\z{-\rec{2mT_f}\z{\v p'-m\overline{\v v}'\z{\v r',t_f}}^2} .
\end{equation}

\subsection{Single particle spectrum}\label{ss:spectrum}

As already mentioned, the calculations leading to the results below are simple Gaussian integrals, as seen from \Eq{e:Srexpr}: the particle number density
$n$ is of Gaussian form, and the Maxwellian term for a velocity field linear in the coordinates also yields a Gaussian shape. The only complication compared
to Refs.~\cite{Csorgo:2001xm,Csorgo:2015scx} is that the desired integrals contain also off-diagonal, $r'_xr'_z$ terms. After some calculation, one arrives
at the following expression for $\td{n}{^3\v p}$ from \Eqs{e:Sdef}{e:N1pdef}:
\begin{equation}
\td{n}{^3\v p'} \propto \exp\z{-\rec{2m}p'_k\z{\v T'}^{-1}_{kl}p'_l},\quad k,l=x,y,z,\label{e:N1expr}
\end{equation}
with summation understood over repeated indices.
We introduced here the $\v T'_{kl}$ matrix and its inverse, ${\v T'}^{-1}_{kl}$, as the matrix whose
components correspond to the inverse slope parameters of the spectrum in the $\overline K'$ frame. We find that the expression of these components is
\begin{align}
T'_{xx} &= T+m\big(\dot X^2 + \omega^2R^2\big),\label{e:Txx}\\ 
T'_{yy} &= T+m\dot Y^2,\\ 
T'_{zz} &= T+m\big(\dot Z^2 + \omega^2R^2\big),\\ 
T'_{xz} &= m\omega R\big(\dot X-\dot Z\big),\label{e:Txz}
\end{align}
and for the inverse matrix:
\begin{equation}\label{e:Tpinvdef}
\begin{pmatrix} T'_{xx} & T'_{xz} \\ T'_{xz} & T'_{zz} \end{pmatrix}
\begin{pmatrix} {T'}^{-1}_{xx} & {T'}^{-1}_{xz} \\ {T'}^{-1}_{xz} & {T'}^{-1}_{zz} \end{pmatrix} =
\begin{pmatrix} 1 & 0 \\ 0 & 1 \end{pmatrix},
\end{equation}
that is,
\begin{align}
{T'}^{-1}_{xx} &= \frac{T'_{zz}}{T'_{xx}T'_{zz}-{T'}^2_{xz}},\label{e:Tinvxx}\\
{T'}^{-1}_{yy} &= \rec{T'_{yy}},\\
{T'}^{-1}_{zz} &= \frac{T'_{xx}}{T'_{xx}T'_{zz}-{T'}^2_{xz}},\\
{T'}^{-1}_{xz} &= \frac{-T'_{xz}}{T'_{xx}T'_{zz}-{T'}^2_{xz}}.\label{e:Tinvxz}
\end{align}
All time-dependent quantities: the axes $X$, $Y$, $Z$, their time derivatives, the $R$ radius, the temperature $T$, and the ``angular velocity''
$\omega$ from \Eq{e:omegadef} are to be taken at the freeze-out time $t_f$, but for brevity we omit the $f$ index in these and in the
following formulas.

We thus got a simple expression for the momentum distribution, much in the spirit of Ref.~\cite{Csorgo:2001xm}. There it was emphasized that for a tilted,
but not rotating ellipsoidal source, the momentum distribution is diagonal in exactly the same frame that corresponds to the tilted ellipsoid. We see that
in our more realistic and general, rotating case, this is clearly not true: the presence of the cross-term in \Eq{e:N1expr} signifes that the eigenframe of
$N_1\z{\v p'}$ is not that of the coordinate-space ellipsoid, $K'$. We may introduce the angle $\vartheta'_\v p$ that corresponds to the tilt of the
eigenframe of the single particle spectrum with respect to the rotated ellipsoids, $K'$. This is given by
\begin{equation}\label{e:thetap}
\tan\z{2\vartheta'_\v p} = \frac{2T'_{xz}}{T'_{xx}-T'_{zz}} = \frac{2\omega R}{\dot X+\dot Z} .
\end{equation}
Thus the observable tilt angle of the momentum spectrum, which we may denote by $\vartheta_\v p$, becomes $\vartheta_\v p \equiv \vartheta+\vartheta'_\v p$.
We see that the additional tilt $\vartheta'_\v p$ of the momentum spectrum depends on how strong is the angular velocity $\omega$ that charactizes the
rotation of the fireball geometry, as compared to an avarege radial Hubble flow, $\frac{\dot X +\dot Z}{2R}$. It is also maybe interesting that in
our model, this $\vartheta'_\v p$ angle does not depend on the particle mass $m$; this gives an indication that in a Buda-Lund model type of relativistic
extension (see the discussion before Section~\ref{ss:sourcefunction}) this angle will not depend on $m_t$.

We can also write up the momentum distribution in the original, laboratory frame ($K$) by the use of the matrix $\v M$ that connects $\v p'$ with $\v p$ in
\Eq{e:ppcomma}:
\begin{equation}\label{e:N1p}
\td{n}{^3\v p} \propto \exp\z{-\rec{2m}p_k\z{\v T}^{-1}_{kl}p_l},\quad k,l=x,y,z
\end{equation}
where
\begin{equation}\label{e:TinvM}
\v T^{-1} = \v M^{-1}{\v T'}^{-1}\v M.
\end{equation}
Written up in components, the inverse slope parameters of the single-particle spectrum in the $K$ frame are
\begin{align}
T^{-1}_{xx} &= {T'}^{-1}_{xx}\cos^2\vartheta+{T'}^{-1}_{zz}\sin^2\vartheta + {T'}^{-1}_{xz}\sin(2\vartheta) , \\
T^{-1}_{yy} &= {T'}^{-1}_{yy},\\
T^{-1}_{zz} &= {T'}^{-1}_{xx}\sin^2\vartheta+{T'}^{-1}_{zz}\cos^2\vartheta - {T'}^{-1}_{xz}\sin(2\vartheta) , \\
T^{-1}_{xz} &= {T'}^{-1}_{xz}\cos(2\vartheta) + \big({T'}^{-1}_{zz}-{T'}^{-1}_{xx}\big)\cos\vartheta\sin\vartheta .
\end{align}
The difference to the formulas in Ref.~\cite{Csorgo:2001xm} is again that the ``intrinsic'' cross-term, ${T'}^{-1}_{xz}$, does appear here: even in the
$\vartheta_f = 0$ (hypothetical) case, one would get cross-terms in the single-particle spectrum.

\subsection{Azimuthal anisotropies}\label{ss:vn}

We can also calculate the azimuthal dependence of the particle production, which is usually characterized by the $v_n$ azimuthal harmonics. Remember, the
$z$ axis is taken as the axis of collision of the nuclei, and the $x$ axis was taken to be the collision event plane, so the $\varphi$ azimuthal angle is
measured in the $x$--$y$ plane. The definition of the azimuth-averaged single-particle spectrum $\td{n}{p_t\m dy}$ and that of the $v_n$, $n=1,2,\dots$
anisotropy parameters is
\begin{equation}\label{e:N1Teff}
\td{n}{^3\v p} = \frac{E}{2\pi p_t}\td{n}{p_t\m dy}\Bigg[1+2\sum_{n=1}^\infty v_n\cos\sz{n\z{\varphi-\Psi_n}}\Bigg] ,
\end{equation}
where $y = \rec 2\ln\frac{E+p_z}{E-p_z}$ is the rapidity, $p_t = \sqrt{p_x^2+p_y^2}$ is the transverse momentum, and $\Psi_n$ is called the $n$th order
event plane angle. 

The calculation of the $v_n$ parameters and the angle-averaged spectrum in our case goes very much similarly if not identically to that found in
Ref.~\cite{Csorgo:2001xm}. It was pointed out there that the angle-averaged spectrum as well as the $v_n$ parameters depend on the kinematical
variables only through certain combinations of them. It is the case also here; the difference is that the expression of these combinations
differ from the earlier results because of the presence of rotation in the velocity terms.

Introducing the $w$ and $v$ variables and the average slope parameter $T_{\rm eff}$ as
\begin{equation}
T_{\rm eff} \equiv \frac{2}{T^{-1}_{xx}+T^{-1}_{yy}},
\end{equation}
\begin{equation}\label{e:vwdef}
w \equiv \frac{p_t^2}{4m}\z{T^{-1}_{xx}-T^{-1}_{yy}},\quad v \equiv-\frac{p_tp_z}{m}T^{-1}_{xz},
\end{equation}
the single-particle spectrum can be written as
\begin{equation}
\td{n}{^3\v p}\propto\exp\z{-\fracd{p_z^2}{2m T_{zz}}
			-\fracd{p_t^2}{2mT_\m{eff}}}\times e^{w{\cos}(2\varphi)+v{\cos}\varphi}.\label{e:dndpvw}
\end{equation}
We can proceed from \Eq{e:dndpvw} by expanding the $v$-dependence in a series. Then using the $I_\nu(w) \equiv \rec\pi\int_0^\pi\m d\varphi
\cos\z{\nu\varphi}e^{w\cos\varphi}$ modified Bessel functions, we can do the Fourier decomposition in the $\varphi$ dependence. It is important to note that
in our simple model all the event planes coincide and this plane is where we set the zero of the azimuthal angle $\varphi$. We obtain the following:
\begin{equation}\label{e:dndptdy}
\frac{\m dn}{2\pi p_t\m dp_t\m dy}\propto \exp\bigg(
-\frac{p_z^2}{2m}T^{-1}_{zz}
-\frac{p_t^2}{2mT_\m{eff}}\bigg)\c I_0\z{w,v},
\end{equation}
\begin{equation}\label{e:vnexpr}
v_n = \frac{\c I_n\z{w,v}}{\c I_0\z{w,v}},
\end{equation}
where the $\c I_n\z{w,v}$ auxiliary quantities are expressed as
\begin{align}
\c I_{2p}\z{w,v}  &\equiv  \sum_{k,l}\frac{I_{\ae{k+p}}(w)+I_{\ae{k-p}}(w)}{2^{2k+2l+1}\z{2k+l}!l!}v^{2k+2l},\\
\c I_{2p+1}\z{w,v}&\equiv v\sum_{k,l}\frac{I_{\ae{k+p}}(w)+I_{\ae{k-p}}(w)}{2^{2k+2l}\z{2k+l}!\z{l+1}!}v^{2k+2l}.
\end{align}
In this latter formula, the summation over $k$ and $l$ formally goes over all integer values of them (including negative ones),
but because of the factorials in the denominator, many terms will be zero.

We have given the full $v$-dependent expansion here. Practically, at mid-rapidity (i.e. arount $p_z = 0$) $v\approx 0$, so only the first few terms in
$v$ are of interest. For the $v_1$ (directed flow), the $v_2$ (elliptic flow) and $v_3$ (third flow), these approximate expressions are:
\begin{equation}\label{e:I0expr}
\c I_0\z{w,v} = I_0(w) + \frac{v^2}{4}\sz{I_0(w) + I_1(w)} + \c O\z{v^4},
\end{equation}
\begin{align}
&v_1 = \frac v2\sz{1+\frac{I_1(w)}{I_0(w)}}+\c O\z{v^3},&\\
&v_2 = \frac{I_1(w)}{I_0(w)} + \frac{v^2}{8}\sz{1+\frac{I_2(w)}{I_0(w)}-2\frac{I_1^2(w)}{I_0^2(w)}}+\c O\z{v^4},&\\
&v_3 = \frac v2\frac{I_2(w)+I_1(w)}{I_0(w)}+\c O\z{v^3}.&
\end{align}
The formulas obtained here are one-to-one copies of those found in Ref.~\cite{Csorgo:2001xm}\footnote{We retained the notation $v$ for the scaling variable
introduced in \Eq{e:vwdef} as a similar quantity appeared in Ref.~\cite{Csorgo:2001xm}; it should not be confused neither with the flow quantities
$v_1$, $v_2$, $v_3$\dots, nor with the fluid velocity $\v v$. Also, the $w$ scaling variable should not be confused with the $\omega$ angular velocity
parameter.}. The difference, as mentioned already, lies in the fact that the relation between
the $w$, $v$ scaling variables and the fundamental kinematical quantities is different here (because the $\v T^{-1}$ matrix contains the effect
of rotation, not only the finite tilt angle). In particular, the apperarance of the non-zero cross term ${T'}^{-1}_{xz}$
implies that $T^{-1}_{xz}$ is nonzero even in the hypothetical case of zero $\vartheta_f$ tilt angle. Thus there is an interplay between the
rotational motion of the fluid and the tilted state of the freeze-out ellipsoids that results in the characteristic rapidity and $p_t$ dependence of the
flow parameters through the $w$ and $v$ variables.

It is important to note that at mid-rapidity $p_z=0$, hence $v=0$, and we recover the simple universal scaling form of the elliptic flow,
$v_2=I_1(w)/I_0(w)$, so the triaxial, rotating and expanding ellipsoids have the same centrality, particle type, collision energy and
transverse momentum independent universal scaling as predicted in Ref.~\cite{Csorgo:2001xm}, and extended to relativistic kinematics in
Ref.~\cite{Csanad:2005gv}. In other words, triaxial ellipsoidal expansion does not spoil the universal scaling of the elliptic flow, but
it modifies the definition of the scaling variable $w$.

\subsection{Two-particle correlations}\label{ss:HBT}

Using the formula \Eq{e:C2def} together with \Eq{e:Sdef}, a straightforward calculation leads to the following expression of the HBT correlation function
in the $K'$ frame (i.e. in the eigenframe of the tilted coordinate-space ellipsoid):
\begin{flalign}\label{e:CKqpexpr}
C\z{\v K',\v q'} = 1+\lambda\exp\z{-\sum_{k,l=x,y,z}q'_k{\v R'}^2_{kl}q'_l}. 
\end{flalign}
Again, the exponent is easier to write down in this matrix form. The components of the ${\v R'}^2$ matrix turn out to be
\begin{align}
{R'}^2_{xx} &= X^2T{T'}^{-1}_{xx},\label{e:Rcommaxx}\\ 
{R'}^2_{yy} &= Y^2T{T'}^{-1}_{yy},\\ 
{R'}^2_{zz} &= Z^2T{T'}^{-1}_{zz},\\ 
{R'}^2_{xz} &= XZT{T'}^{-1}_{xz},\label{e:Rcommaxz}
\end{align}
where the components of the inverse temperature matrix ${\v T'}^{-1}$ are given by \Eqsdash{e:Tinvxx}{e:Tinvxz}. 

The fact that the radius parameters do not depend on the total transverse momentum $\v K$ of the pair is a feature characteristic to the non-relativistic
nature of the treatment and the self-similar nature of the solution~\cite{Csorgo:2001xm}. Again, as before Section~\ref{ss:sourcefunction}, we mention that
in a yet to be explored relativistic generalization, the HBT radii most probably depend on the transverse mass $m_t$ of the pair approximately in the same
way as they do depend on the particle mass $m$ in the case of our exact non-relativistic solution. In our presented case this dependence is rather
involved; it is given by \Eqsdash{e:Rcommaxx}{e:Rcommaxz}, which in turn refer to \Eqs{e:Tinvxx}{e:Tinvxz} and \Eqs{e:Txx}{e:Txz}. However, it is not
hard to see that a term with approximate $1/m$-like dependence is present in the expression of the squared radius parameters, which implies the
well known $1/R^2 = C_1+C_2\cdot m_t$-like dependence of the $R^2$ components in the relativistic setting.
In what follows, however, we again concentrate on the exact non-relativistic results, the relativistic generalization being outside of the
scope of this paper. 

To analyze the obtained HBT correlation further, we might again note that the final eigenframe of the rotating ellipsoid, $K'$ is \emph{not} the same
as the frame in which the (Gaussian-like) HBT correlation function $C\z{\v q'}$ is diagonal. We denote the angle between $K'$ and the eigenframe of the
HBT correlation function by $\vartheta'_\m{HBT}$. It turns out that this angle is not only non-zero, but in general different from $\vartheta'_\v p$,
the angle that described the eigenframe of the single-particle spectrum. The expression of $\vartheta'_\m{HBT}$ is
\begin{eqnarray}
\null & \null & \tan\z{2\vartheta'_\m{HBT}}  =  \frac{2XZT'_{xz}}{X^2T'_{zz}-Z^2T'_{xx}} = \\ \nonumber
\null &\null  & =   \frac{2mXZ\omega R\z{\dot X-\dot Z}}{\z{T+m\omega^2R^2}\z{X^2-Z^2}+m\big(X^2\dot Z^2-Z^2\dot X^2\big)} ,
\label{e:thetaHBT}
\end{eqnarray}
thus the observable tilt angle of the HBT correlation function, which we may denote by $\vartheta_\v q$, becomes $\vartheta_\v q\equiv\vartheta+
\vartheta'_\m{HBT}$. In contrast to the $\vartheta'_\v p$ angle introduced in \Eq{e:thetap}, this $\vartheta'_\m{HBT}$ angle does depend on the mass of the
particle, thus in a relativistic setting it may pick up an $m_t$-dependence. It might be interesting to note that by formally setting $m = 0$ in the above
formula, the $\vartheta'_\m{HBT}$ angle vanishes. Thus in the vanishing transverse mass limit the measurable tilt of the HBT system, $\vartheta_\v q$
approaches the actual $\vartheta$ angle of tilt of the geometrical shape of the triaxial ellipsoid of the expanding and rotating fireball. The latter
angle, denoted by $\vartheta$ up until now, may in this context thus be denoted by $\vartheta_\v r$, being the geometrical tilt.

To write up the HBT correlation function in the laboratory frame ($K$ frame), we only need to apply the matrix $\v M$ introduced in \Eq{e:Mv} to the
components of $\v q'$ to express $\v q'$ with the components of the relative momentum measured in the $K$ frame. Simple calculation leads to
\begin{equation}\label{e:CKqexpr}
C\z{\v K,\v q} = 1+\lambda\exp\Bigg(-\sum_{k,l=x,y,z}q_k\v R^2_{kl}q_l\Bigg), 
\end{equation}
\begin{align}
R^2_{xx} &= {R'}^2_{xx}\cos^2\vartheta + {R'}^2_{zz}\sin^2\vartheta + {R'}^2_{xz}\sin\z{2\vartheta},\\ 
R^2_{yy} &= {R'}^2_{yy},\\ 
R^2_{zz} &= {R'}^2_{xx}\sin^2\vartheta + {R'}^2_{zz}\cos^2\vartheta - {R'}^2_{xz}\sin\z{2\vartheta},\\ 
R^2_{xz} &= {R'}^2_{xz}\cos\z{2\vartheta} + \big({R'}^2_{zz}-{R'}^2_{xx}\big)\sin\vartheta\cos\vartheta.
\end{align}
We can also evaluate the HBT radius parameters suited for the usual setting of azimuthally sensitive HBT measurements, in the so-called Bertsch-Pratt
(BP) or out-side-long frame. In this frame, the relative momentum vector $\v q$ is written up in the components $\z{q_\m{l},q_\m{o},q_\m{s}}$:
the $q_\m{l}$ (``long'') component points in the beam (that is, the $z$) direction, the $q_\m{o}$ (``out'') component points to the direction
of $\v K$, the average transverse momentum of the pair, and $q_\m{s}$ (``side'') is the component perpendicular to both of these. We denote
the azimuthal angle of $\v K$ in the $x$--$y$ plane by $\varphi$, so
\begin{align}
q_\m{l} &= q_z,\\
q_\m{o} &=  q_x\cos\varphi + q_y\sin\varphi,\\
q_\m{s} &= -q_x\sin\varphi + q_y\cos\varphi.
\end{align}
An important additional remark is in order here (just as in Ref.~\cite{Csorgo:2001xm}): the preceding formulas were derived for instantaneous
particle emission at time $t_f$. Assuming a finite $\Delta t$ time duration of the particle emission (eg. by setting the time dependence as
$(2\pi\Delta t^2)^{-1/2}\exp[-(t-t_f)^2/ 2\Delta t^2]$, a Gaussian) will have an effect on the HBT correlation function (although not on the single-particle
spectrum). As in Refs.~\cite{Csorgo:2001xm,Csorgo:2015scx}, one gets the result that the radius parameters have to be augmented with an additional term
$\delta R^2_{ij} = \beta_i\beta_j \Delta t^2$, where $\gvec\beta =({\v p_1+\v p_2})/({E_1 + E_2})$ is the velocity of the pair. In the Bertsch-Pratt frame
$\beta_\m{s} = 0$, so finally we have the correlation function as 
\begin{equation}\label{e:CBPexpr}
C\z{\v K',\v q'} = 1+\lambda\exp\Bigg(-\sum_{k,l=\m o,\m s,\m l}q_k\v R^2_{kl}q_l\Bigg), 
\end{equation}
with the Bertsch-Pratt radius parameters being equal to
\begin{align}
R^2_\m{oo} &= R^2_{xx}\cos^2\varphi + R^2_{yy}\sin^2\varphi+\beta^2_\m{o}\Delta t^2,\label{e:BPoo}\\
R^2_\m{ss} &= R^2_{xx}\sin^2\varphi + R^2_{yy}\cos^2\varphi,\\
R^2_\m{ll} &= R^2_{zz}+\beta^2_\m{l}\Delta t^2,\\
R^2_\m{os} &= \z{R^2_{yy}-R^2_{xx}}\sin\varphi\cos\varphi,\\
R^2_\m{ol} &= R^2_{xz}\cos\varphi+\beta_\m{l}\beta_\m{o}\Delta t^2,\\
R^2_\m{sl} &= -R^2_{xz}\sin\varphi.\label{e:BPsl}
\end{align}
Note that in the Longitudinally Co-Moving System, (LCMS), the above formulas simplify as in this system $\beta_\m{l}=0$. In particular,
\begin{align}
R_\m{ll}^2 & = R_{zz}^2,\\
R_\m{ol}^2 & = R_{xz}^2 \cos(\varphi),
\end{align}
hence in the LCMS we obtain the interesting relation:
\begin{equation}\label{e:Rolsl}
R_\m{ol}^2(\varphi+\pi/2) = R_\m{sl}^2(\varphi),
\end{equation}
and both terms oscillate in $\varphi$ with half of the frequency of the oscillations of the side--side and out--out terms and the out--side cross-term.
This feature may be straightforward to test experimentally.

We thus see how the oscillation of the Bertsch-Pratt radii (especially of the
out-long, side-long components) is connected to the rotation of the flow. We also see evidently the emergence of a long-known fact that the
$\varphi$-averaged $R^2_\m{oo}-R^2_\m{ss}$ value (at a given $m$ value) basically measures the duration of the particle emission: this is a frequently
exploited feature, perhaps most recently in the already mentioned work of Ref.~\cite{Lacey:2014wqa}, where the non-monotonic behavior
and finite-size scaling properties of this quantity is used to give a first indication on the presence of a critical endpoint on the phase diagram of QCD.

As an aside: note, however, that for expanding fire-shells with large temperature inhomogeneity and relatively small radial flows, typical for hadron-proton
or proton-proton collisions, $R_\m{ss}<R_\m{oo}$ is also possible and actually expected at low transverse momentum, as predicted in Ref.~\cite{Csorgo:1995bi}
and as indicated as a robust feature of h+p and p+p reactions in Refs.~\cite{Agababyan:1997wd,Csorgo:1999sj,Bialas:2014boa}.

\section{Illustration of results and discussion}\label{s:discussion}

We have demonstrated in the previous section how the rotating nature of the flow of our presented solution translates to the observable quantities.
In this section we illustrate the results obtained above by taking a reasonable set of initial conditions.

In the hydrodynamical equations we set the $m_0$ mass to be the proton mass, $m_0 = 938$ MeV, and start the time evolution with $T_0 = 300$ MeV. 
As an illustration, we take the initial conditions for the principal axes as $X_0 = 4$ fm, $Y_0 = 6$ fm, $Z_0 = 2$ fm, the initial ellipsoid being the
thinnest in the beam direction, resembling the conditions right after a non-central heavy-ion collision. The $\omega_0$ parameter is taken to be $0.15$
$c$/fm. As said earlier, we make the assumption that the freeze-out happens instantaneously when the temperature reaches $T_f$, taken to be 140 MeV. 
(The parameter set employed here closely follows the one taken in Ref.~\cite{Csernai:2014hva} where the spheroidal special case of our solution was studied.)

One of our goals in this section is to demonstrate that the final state observables carry information on the rotation and thus on the equation of state.
In this work we do not investigate all the possible equations of state (including the $\mu_B=0$ lattice QCD EoS, and those calculated for finite $\mu_B$)
in detail, we just want to give a hint at how the softness of the equation of state might influence the time evolution and the final state observables,
deferring the detailed investigation to a follow-up work. Here we simply take different constant values of $\kappa$ in the equation of state, \Eq{e:eos}.
The conclusions that we arrive at with the initial conditions and assumptions should not thus be taken as general conclusions. But nevertheless, we will
be able to draw some qualitative conclusions, and we will hint at which of the features of our results may carry over to a more general setting.

\begin{figure}%
\begin{center}
\ifdefined\INCLUDEFIGS
\includegraphics[width=0.9\columnwidth]{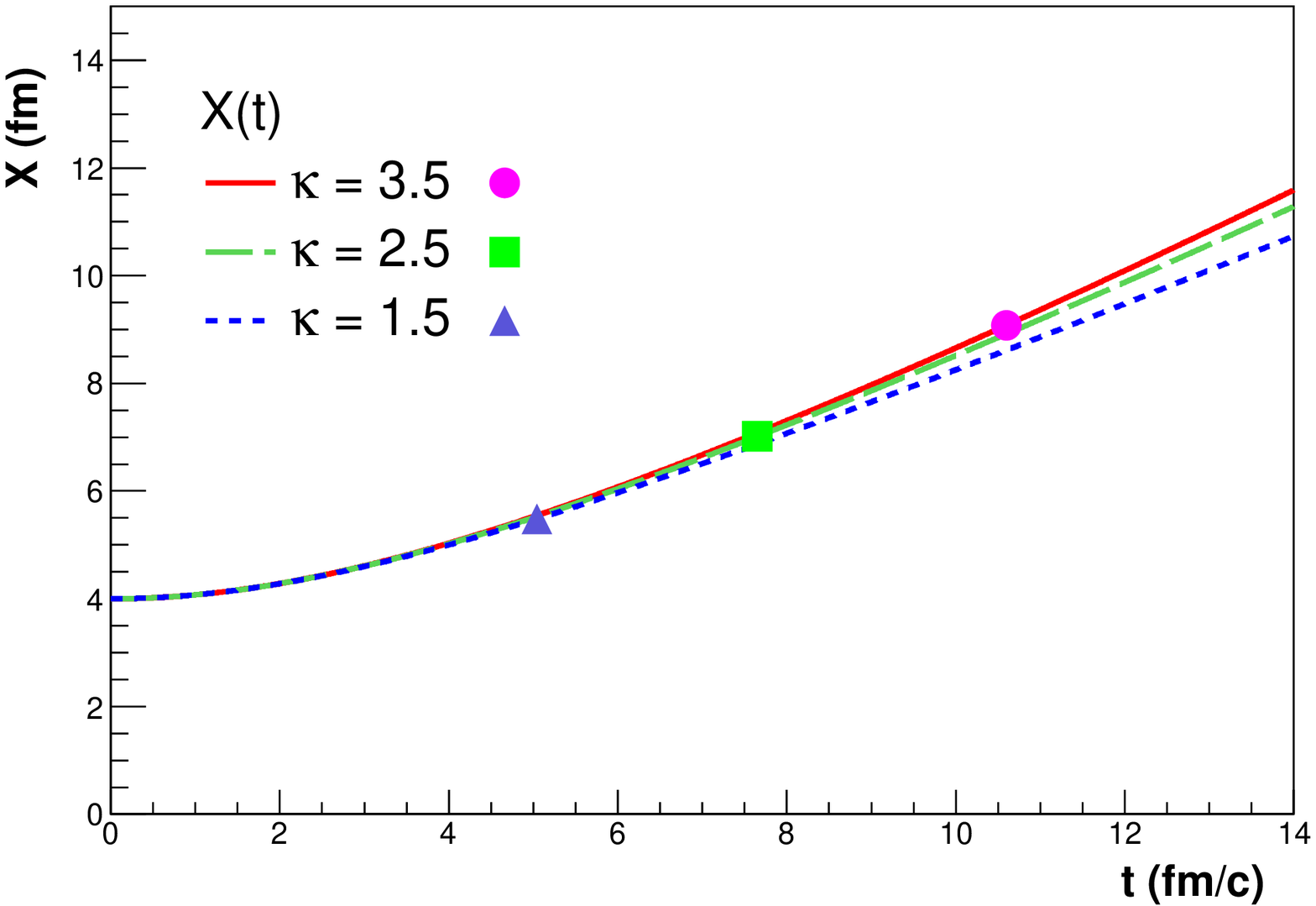}\\
\includegraphics[width=0.9\columnwidth]{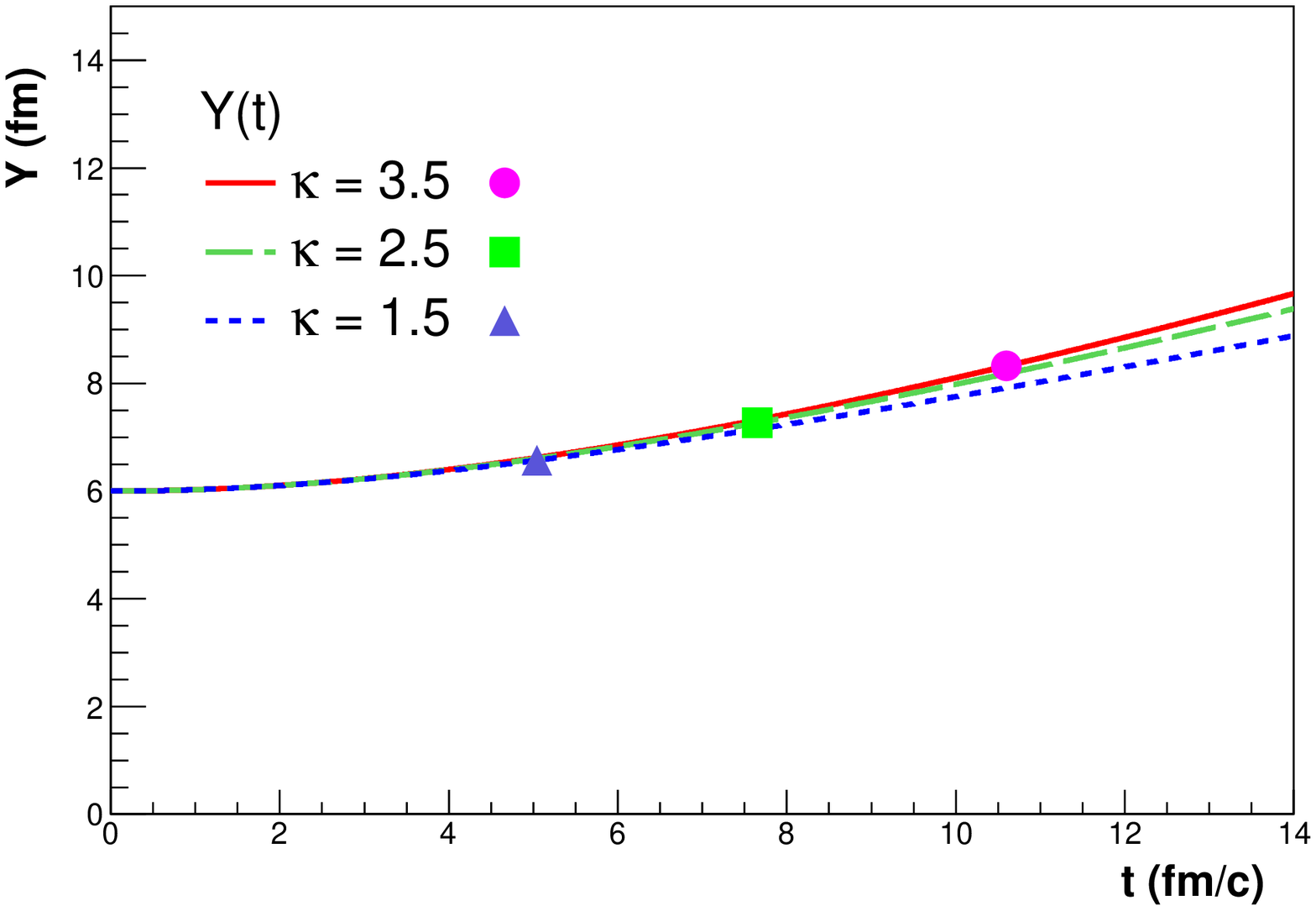}\\
\includegraphics[width=0.9\columnwidth]{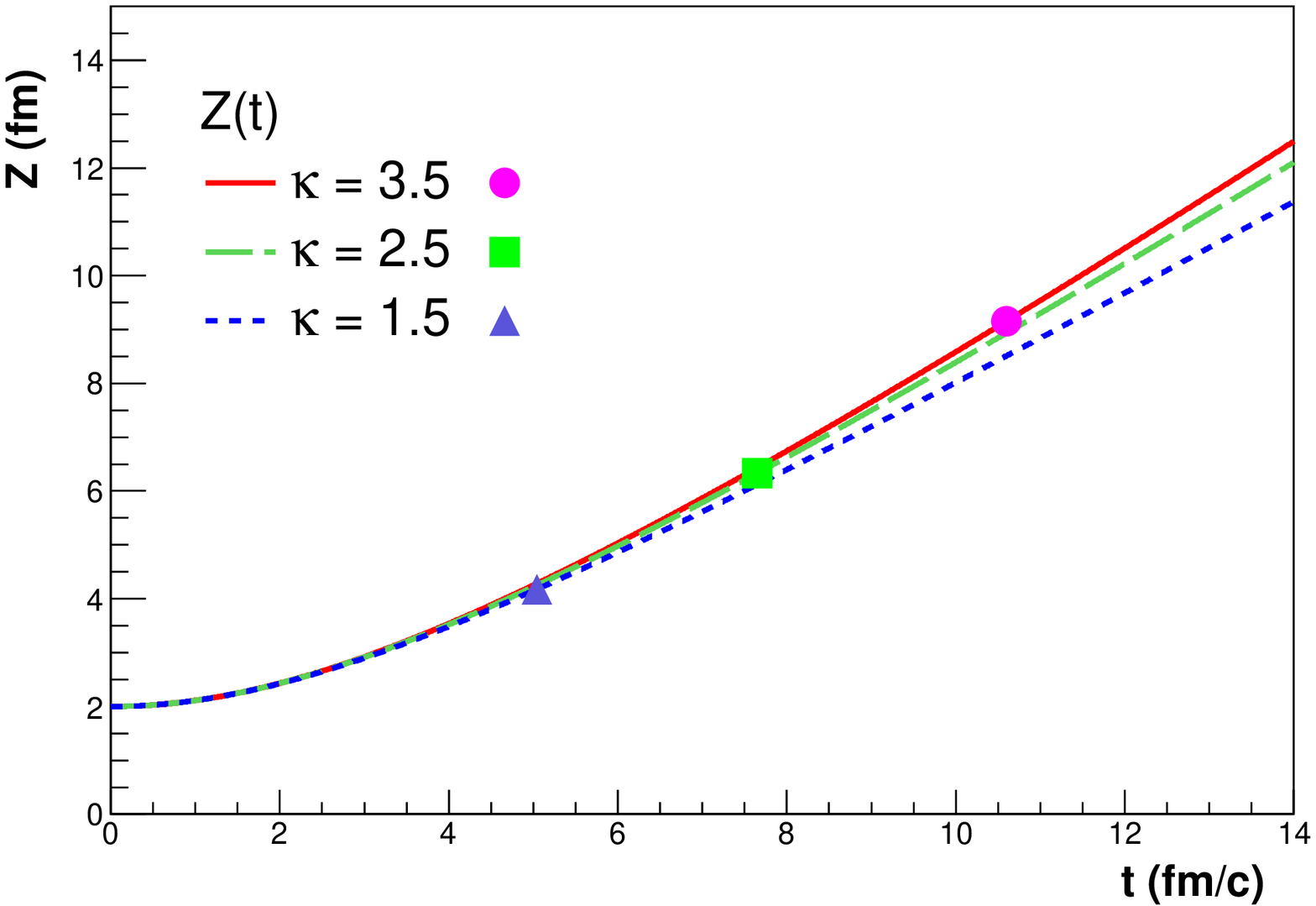}\\
\fi
\caption{Illustration of the dependence of the time evolution of principal axes $X$, $Y$, $Z$ of the solution on the equation of state, for
three different constant $\kappa$ values. Initial conditions and parameters are: $m_0 = 938$ MeV, $T_0 = 300$ MeV, $\omega_0 = 0.15$ $c/$fm, $X_0 = 4$
fm, $Y_0 = 6$ fm, $Z_0 = 2$ fm, and $\dot X_0 = \dot Y_0 = \dot Z_0 = 0$. Markers denote the values at the respective freeze-outs (when the temperature
reaches $T_f = 140$ MeV).}
\label{f:XYZ}%
\end{center}
\end{figure}

\begin{figure}%
\begin{center}
\ifdefined\INCLUDEFIGS
\includegraphics[width=0.9\columnwidth]{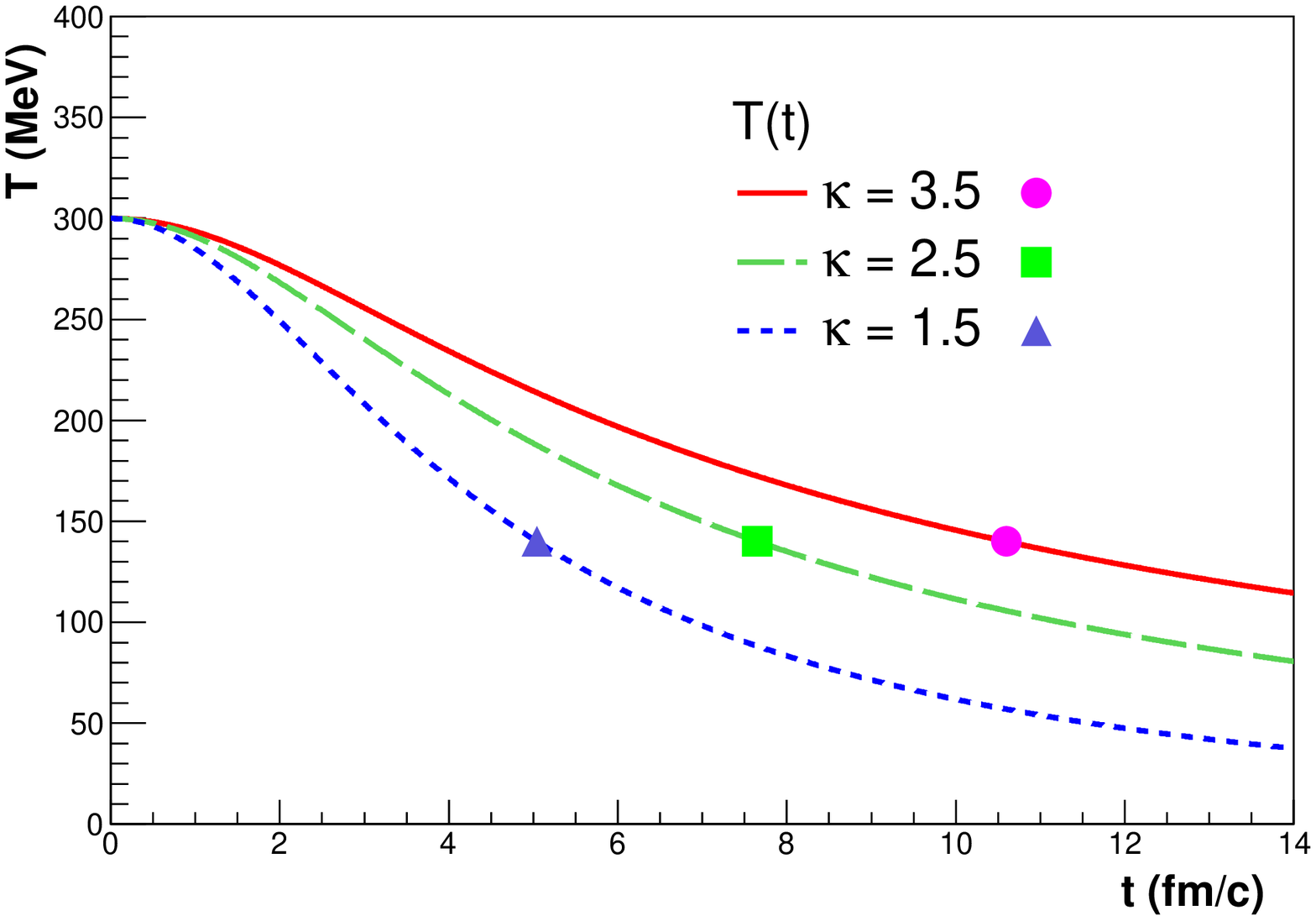}\\
\includegraphics[width=0.9\columnwidth]{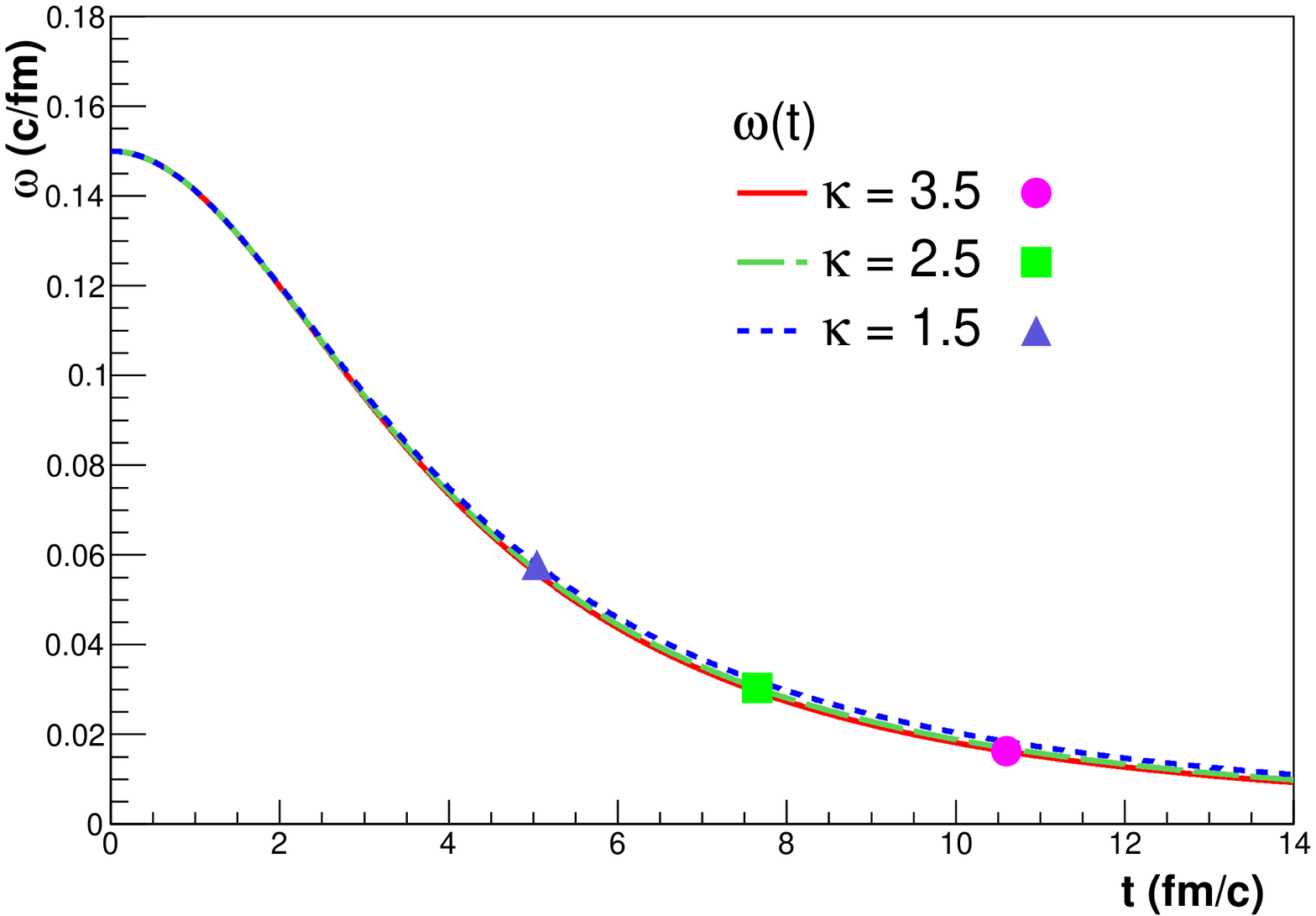}
\fi
\caption{Illustration of the dependence of the time evolution oft the temperature $T$ (upper panel) and the angular velocity $\omega$ introduced in
\Eq{e:omegadef} on the equation of state, for three different constant $\kappa$ values. Initial conditions and parameters are as in the previous example:
$m_0 = 938$ MeV, $T_0 = 300$ MeV, $\omega_0 = 0.15$ $c/$fm, $X_0 = 4$ fm, $Y_0 = 6$ fm, $Z_0 = 2$ fm, and $\dot X_0 = \dot Y_0 = \dot Z_0 = 0$. Markers
denote the values at the respective freeze-outs (when the temperature reaches $T_f = 140$ MeV).}
\label{f:Tomega}%
\end{center}
\end{figure}

On Fig.~\ref{f:XYZ} we plot the time evolution of the principal axes $X(t)$, $Y(t)$, $Z(t)$ of our solution for the mentioned initial conditions and
parameters, for three different $\kappa$ values. The higher the $\kappa$, the ``softer'' the equation of state is. On Fig.~\ref{f:Tomega} we plot also the
time evolution of the temperature $T(t)$ as well as that of the angular velocity $\omega(t)$. We denote the values of these quantities at the freeze-out
by distinct markers. 

\begin{figure}%
\begin{center}
\ifdefined\INCLUDEFIGS
\includegraphics[width=0.9\columnwidth]{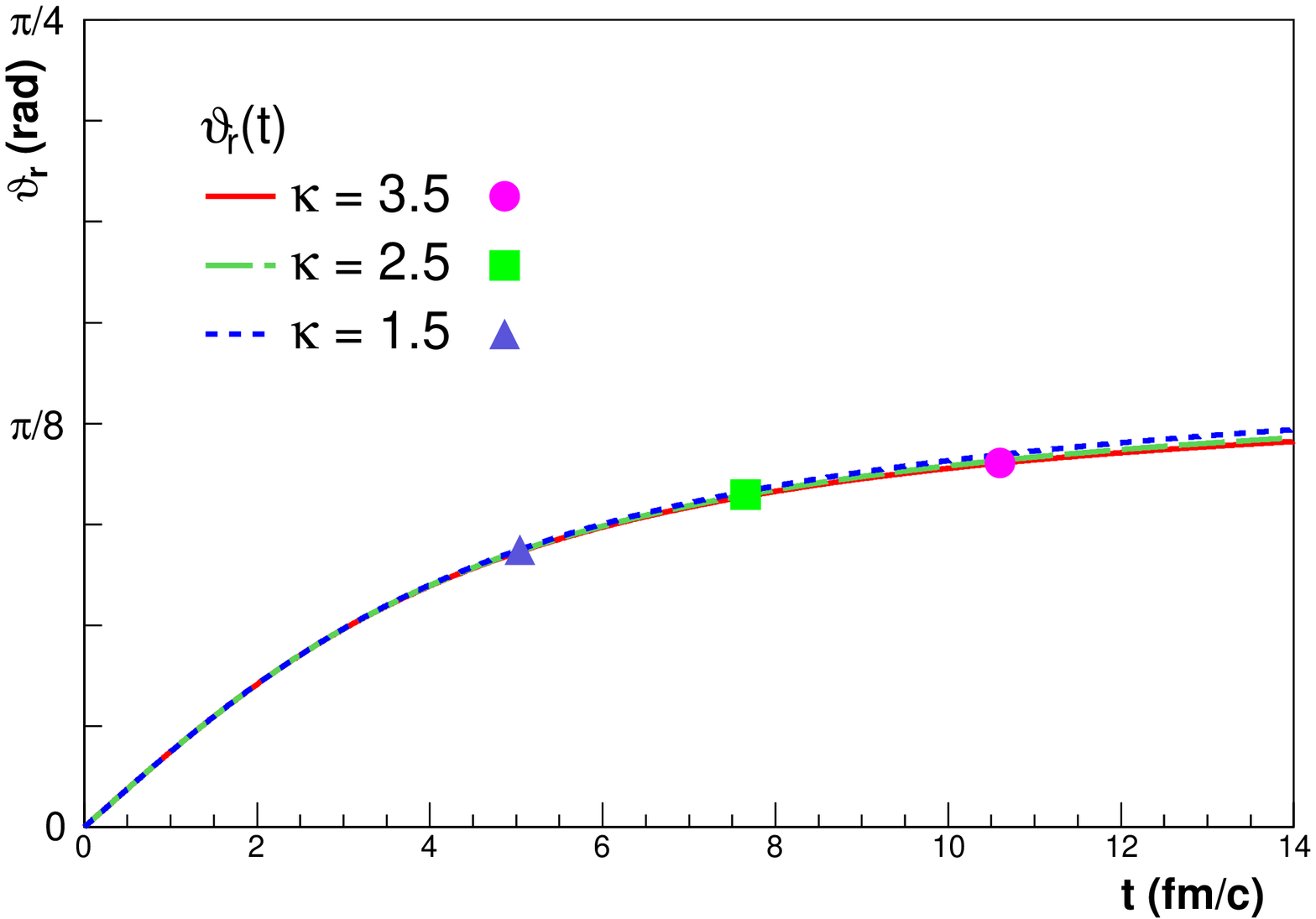}\\
\includegraphics[width=0.9\columnwidth]{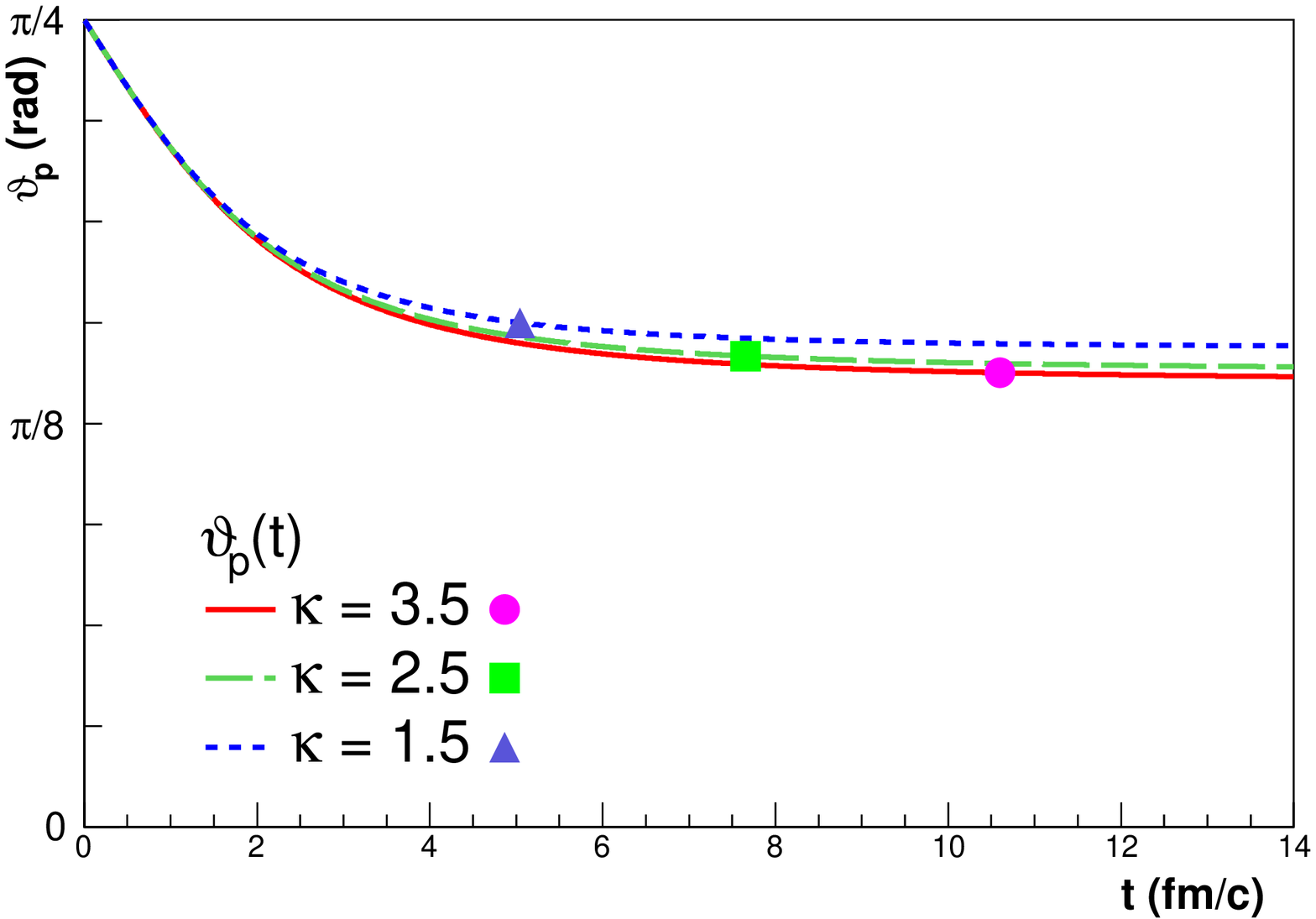}\\
\includegraphics[width=0.9\columnwidth]{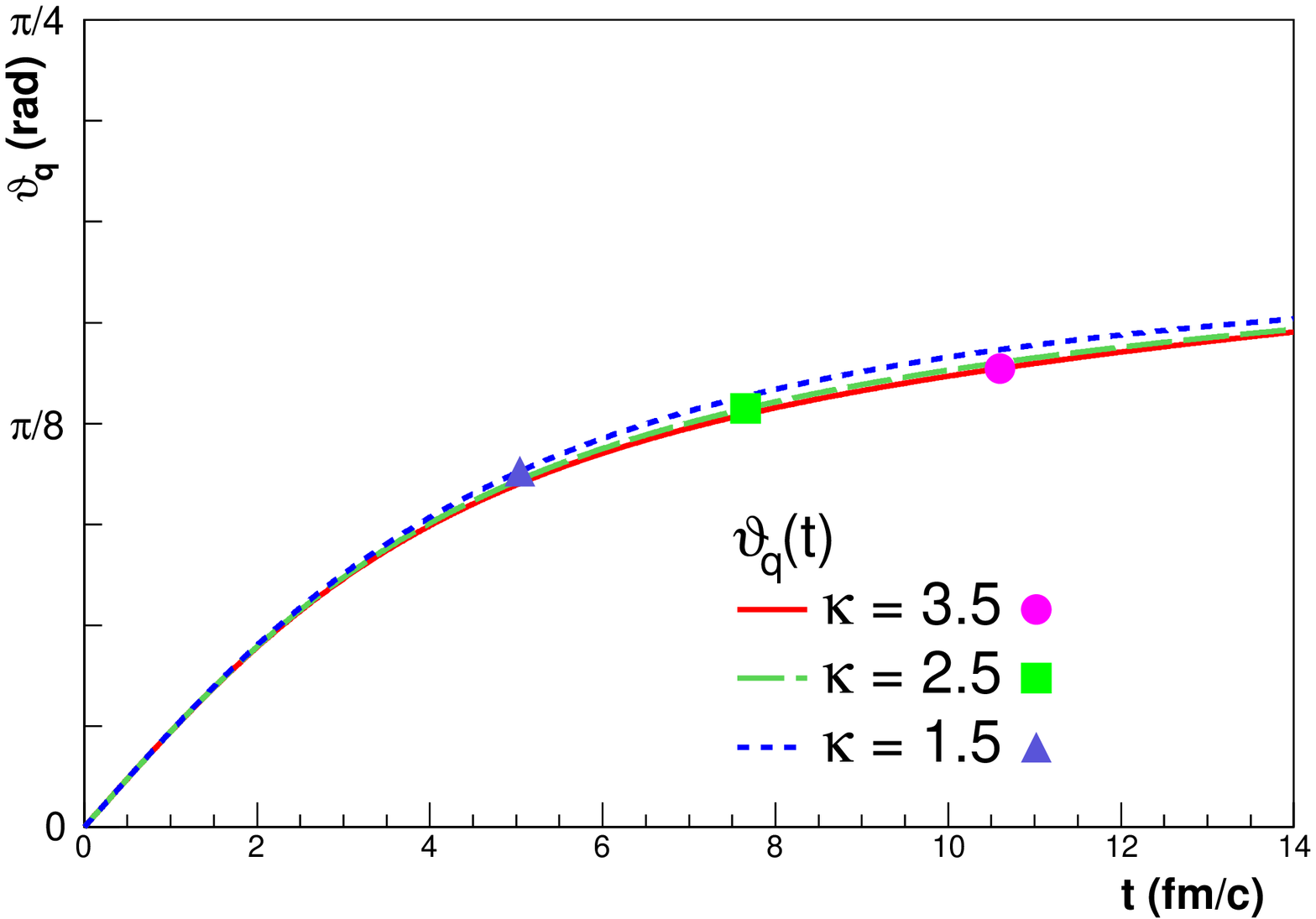}
\fi
\caption{Time evolution of the various tilt angles in our new hydrodynamical solution, for three different EoS. Parameters and initial conditions are
the same as in Figs.~\ref{f:XYZ} and \ref{f:Tomega}, markers denote the values at the respective freeze-outs. Upper panel: the tilt angle of the
coordinate-space ellipsoids in the $x$--$z$ plane, $\vartheta_\v r$, as introduced in the text. This is the tilt angle of the co-rotating $K'$ frame.
Middle panel: the $\vartheta_\v p$ angle, the observable tilt angle of the eigenframe of the single-particle spectrum ($\vartheta_\v p = \vartheta +
\vartheta'_\v p$). Lower panel: $\vartheta_\v q$, the tilt angle of the eigenframe of the HBT correlation function ($\vartheta_\v q = \vartheta +
\vartheta'_\m{HBT}$). For plotting the $\vartheta_\v q$ angle, the mass of the pion $m_\pi$ was used to evaluate $\vartheta'_\m{HBT}$. As mentioned after
\Eq{e:thetaHBT}, in the $m\to 0$ limit, the coordinate-space tilt is recovered as $\vartheta_\v q\to\vartheta_\v r$.}
\label{f:thetaRPH}%
\end{center}
\end{figure}

Further, we plot the time evolution of the various tilt angles of the system introduced so far on Fig.~\ref{f:thetaRPH}. For this sake, we use the unified
notation already introduced: $\vartheta_\v r\equiv \vartheta$ denotes the tilt angle of the coordinate-space ellipsoids (eg. that of the level surfaces of
the particle number density). However, as was pointed out, when rotation of the system plays a role, this angle is not accessible at first hand
experimentally. Rather, what one can measure is the tilt angle corresponding to the eigenframe of the single-particle spectrum, which we can denote by
$\vartheta_\v p \equiv \vartheta+\vartheta'_\v{p}$, where $\vartheta'_\v p$ was introduced in \Eq{e:thetap}, and measures the tilt of the eigenframe
of the momentum spectrum with respect to the $K'$ frame, which itself is tilted by $\vartheta_\v r \equiv \vartheta$ in the laboratory frame. Another
observable tilt angle is that of the eigenframe of the HBT correlation function, which we denote by $\vartheta_\v q = \vartheta+\vartheta'_\m{HBT}$,
where $\vartheta'_\m{HBT}$ measures the tilt angle of the HBT correlation function in the $K'$ frame, and
is introduced in \Eq{e:thetaHBT}. The initial value of these measured angles is sensitive to the precise initial conditions; e.g. the inital value of
$\vartheta_\v p$ of $\pi/4$ is due to the fact that our special initial conditions had $\dot X_0 = \dot Y_0 = \dot Z_0 = 0$.
Nevertheless, one sees that by simultaneously measuring $\vartheta_\v p$ and $\vartheta_\v q$, one can infer the final rotation angle of the system,
and one gets a quantity that is sensitive to the equation of state.

We do not detail further investigations of more specific equation of states now. However, we note that in our plotted case, the EoS dependence of the final
$\vartheta_\v r$ tilt angle mainly comes from the fact that the adiabatic expansion lasts longer for a softer (i.e. that with a higher $\kappa$) equation of
state. In the plotted case, i.e. when $T_0$ is kept fixed as $\kappa$ changes, the change in the time evolution of the principal axes because of the change
in $\kappa$ has an opposite effect: we see from Figs.~\ref{f:XYZ} and \ref{f:Tomega} that in this case, for softer $\kappa$ the system expands \emph{more}
violently. In the plotted case the first effect dominates, so at the end of the day, softer $\kappa$ will result in greater final tilt angle.
It must be noted that if e.g. the total energy density is held fixed as $\kappa$ changes, one can have a different conclusion, since in this case the
time evolution will be \emph{less} violent for softer $\kappa$. Since we do not have any a priori knowledge of the initial temperature or the initial
energy density of the thermalized matter produced in various energy nucleus-nucleus collisions, in a realistic setting, the conclusions evident on
Figs.~\ref{f:XYZ} and \ref{f:Tomega} may undergo significant changes. However, it is clear that besides the final (freeze-out) value of the $\omega(t)$
angular velocity, the final rotation angle of the system is a sensitive additional tool to investigate in the quest for the experimental equation of
state of the strongly interacting matter.

\begin{figure}%
\begin{center}
\ifdefined\INCLUDEFIGS
\includegraphics[width=0.9\columnwidth]{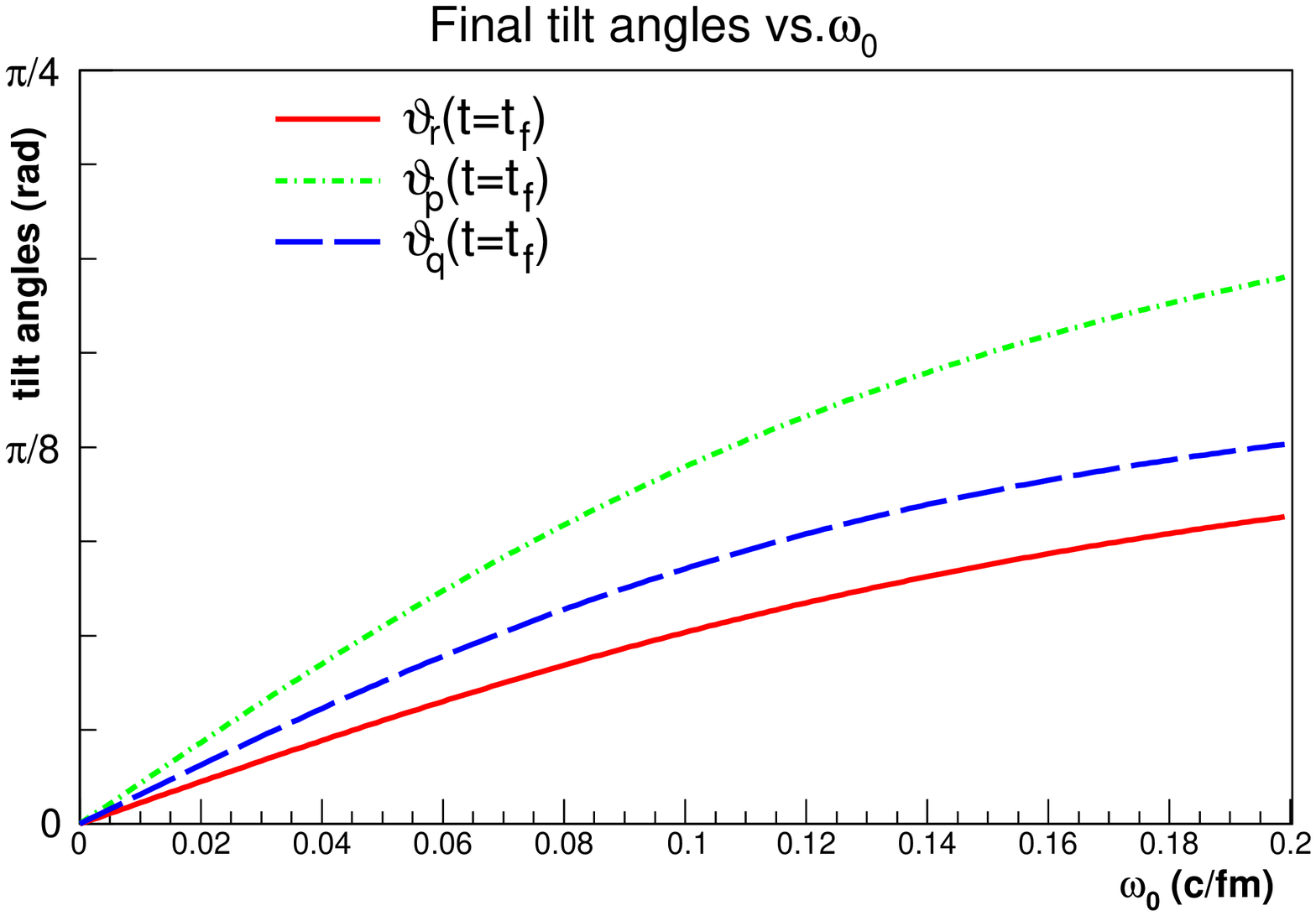}\\
\includegraphics[width=0.9\columnwidth]{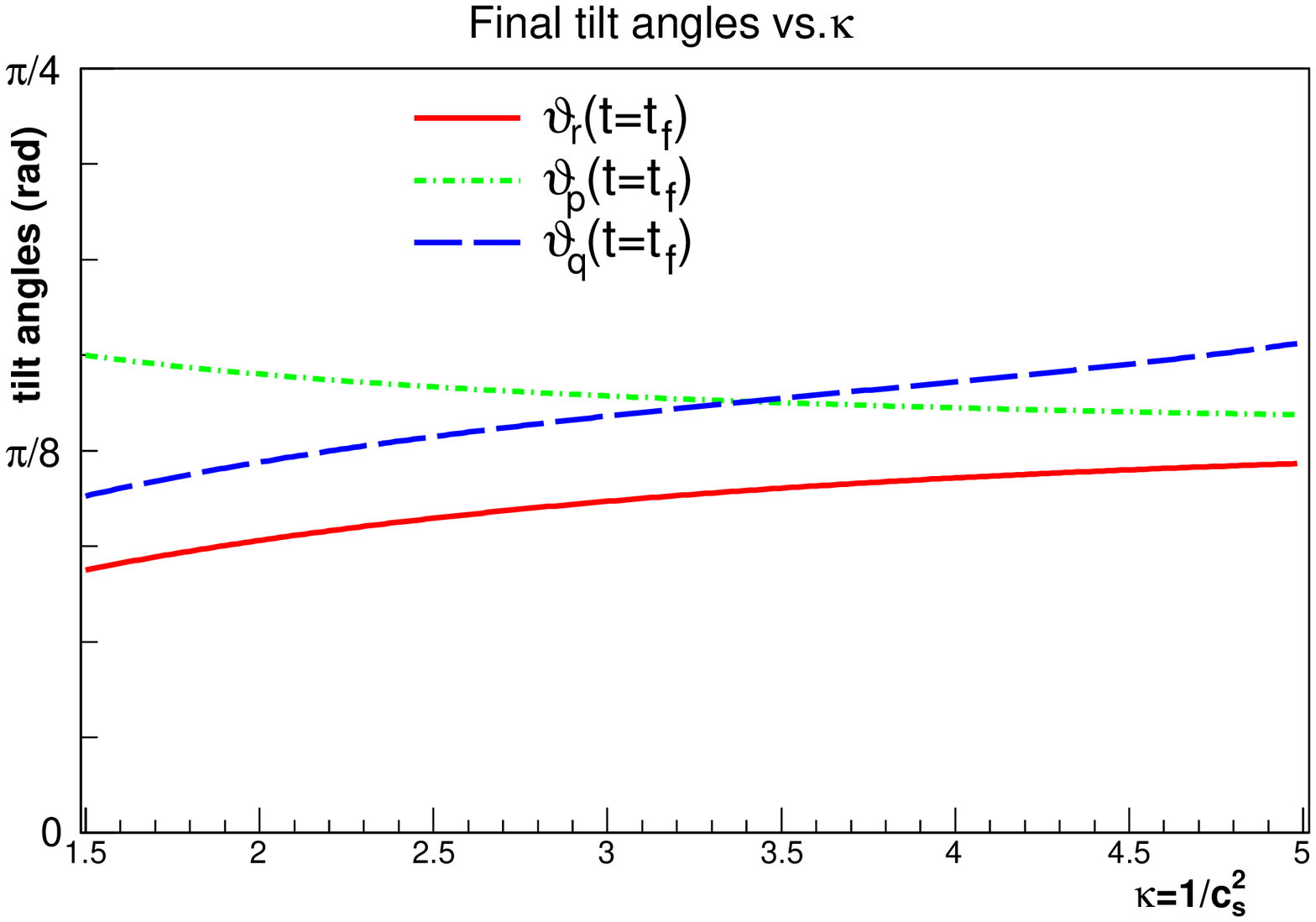}
\fi
\caption{The various tilt angles introduced in the text ($\vartheta_\v r\equiv \vartheta$, the coordinate-space tilt, $\vartheta_\v p\equiv\vartheta +
\vartheta'_\v{p}$, the tilt of the single-particle spectrum, and $\vartheta_\v q \equiv \vartheta+\vartheta'_\m{HBT}$, the tilt of the HBT correlation
function) at freeze-out time. Upper panel: freeze-out time angles plotted as a function of initial angular velocity $\omega_0$ (for $\kappa=3/2$).
Lower panel: freeze-out time angles plotted as a function of $\kappa$ (in this plot, the $\omega_0$ was taken to be $0.15$ $c/$fm).
All other initial conditions are the same as in Fig.~\ref{f:XYZ}.}
\label{f:omegakappa_tR}%
\end{center}
\end{figure}

Fig.~\ref{f:omegakappa_tR} illustrates how the final, freeze-out tilt angles: the $\vartheta_\v r$ (the coordinate-space tilt of the ellipsoid),
and the two observable tilt angles, $\vartheta_\v p \equiv \vartheta+\vartheta'_\v{p}$ (the tilt of the single-particle spectrum), and $\vartheta_\v q \equiv
\vartheta+\vartheta'_\m{HBT}$ (the tilt of the HBT correlation function) depend on the initial condition $\omega_0$, and on the $\kappa$ parameter in
the EoS, respectively. All the other initial conditions and parameters in these plots are the same as those used for Fig.~\ref{f:thetaRPH}.

\begin{figure}%
\begin{center}
\ifdefined\INCLUDEFIGS
\includegraphics[width=0.9\columnwidth]{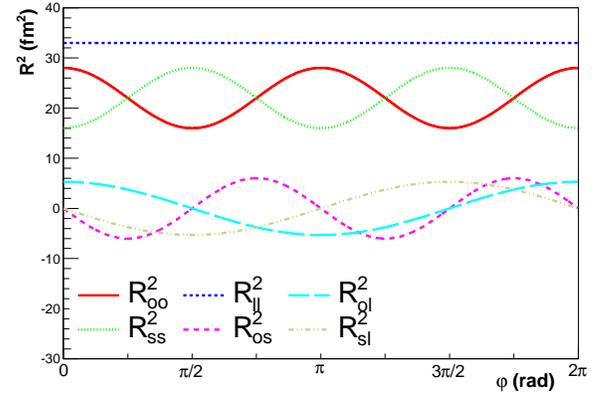}
\fi
\caption{Bertsch-Pratt HBT radii vs. $\varphi$ azimuthal angle of the particle pair, calculated for a reasonable set of parameters: ${R'}^2_{xx} = 25$ fm$^2$,
${R'}^2_{yy} = 16$ fm$^2$, ${R'}^2_{zz} = 36$ fm$^2$, ${R'}^2_{xz} = 2$ fm$^2$, $\vartheta_f = \pi/8$. The
oscillations in $R^2_\m{oo}$, $R^2_\m{ss}$ and $R^2_\m{os}$ with $\pi$ periodicity are characteristic to an ellipsoid-like source. The oscillations in
$R^2_\m{ol}$ and $R^2_\m{sl}$ with $2\pi$ periodicity are characteristic to a tilted or rotating ellipsoid-like source. A measurement of these Bertsch-Pratt
radii enables one to deduce the $\v R$ HBT radius matrix introduced in \Eq{e:CKqexpr}, and in turn the $\vartheta+\vartheta'_\m{HBT}$ angle introduced in
\Eq{e:thetaHBT}, that characterizes the tilt of the eigenframe of the HBT correlation function. In this plot, the freeze-out is assumed to be instantaneous
($\Delta t = 0$), although in real data analysis $\Delta t$ plays an important role.}
\label{f:HBTBP}%
\end{center}
\end{figure}

\begin{figure}%
\begin{center}
\ifdefined\INCLUDEFIGS
\includegraphics[width=0.9\columnwidth]{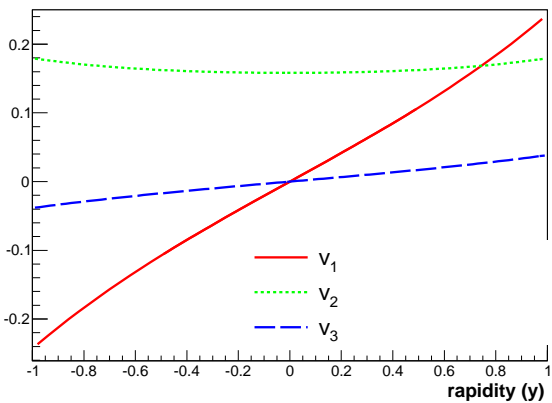}
\fi
\caption{
Illustration of the dependence of flow parameters $v_1$, $v_2$, $v_2$ on rapidity $y$. The parameters were set to ${T'}_{xx} = 300$ MeV, $T'_{xz} = -20$ MeV,
$T'_{zz} = 500$ MeV, $T'_{yy} = 200$ MeV, $\vartheta_f = \pi/8$. The $p_t$ of the particles are assumed to be $p_t = 600$ MeV$/c$, while the particle mass
$m$ was set to equal to the kaon mass, $m=494$ MeV. By measuring these flow coefficients and the angle-averaged spectrum,
one can reconstruct the $\v T$ slope matrix elements introduced in \Eq{e:N1p}, and in turn the angle $\vartheta'_\v{p}$ that characterizes the tilt of the
eigenframe of the momentum distribution, as introduced in \Eq{e:thetap}.
}
\label{f:vn}%
\end{center}
\end{figure}

We plot some usual observable quantities such as the Bertsch-Pratt HBT radii given by \Eqsdash{e:BPoo}{e:BPsl} as a function of pair azimuthal angle on
Fig.~\ref{f:HBTBP}, and the rapidity dependence of the azimuthal harmonics of the single particle spectrum, the $v_1$ directed flow, the $v_2$ elliptic flow,
and the $v_3$ third flow on Fig.~\ref{f:vn}, for a reasonable set of parameter values at freeze-out. The intention of these plots is to illustrate the
behavior of usual observables; we note that a combined
measurement of all the HBT radii, including the $R^2_\m{ol}$, $R^2_\m{sl}$ cross-terms is necessary to determine the $\vartheta'_\m{HBT}$ angle. In particular,
as seen from \Eqsdash{e:BPoo}{e:BPsl}, the out-long and side-long cross-terms are the most characteristic to the tilted ellipsoidal source (as already pointed
out in Refs.~\cite{Lisa:2000ip,Csorgo:2001xm}), but here we see that not only the final tilt angle but the rotational motion also gives a contribution to
these parameters. Experimentally, the measurement of these cross-terms are the most challenging, because one has to have a combined event-by-event information
on the first and second order event planes. In our simple hydrodynamical model, these event planes coincide, but in a realistic setting, both of them will
be smeared by initial state fluctuations. Fig.~\ref{f:HBTBP} also illustrates the \r{e:Rolsl} relation between the $R_\m{ol}$ and $R_\m{sl}$ cross-terms:
that in the LCMS, $R_{ol}^2(\phi+\pi/2)=R_{sl}^2(\varphi)$, and both terms oscillate with half of the frequency of the oscillations of the other, more
commonly measured out--out, side--side and out--side terms.

Also, a combined measurement of at least the slope of the $v_1(y)$, the $p_t$ dependence of $v_2$ and the measurement of the angle-averaged single-particle
spectrum is necessary to get the $\vartheta'_\v{p}$ value. The most characteristic feature stemming from a tilted (or rotating) source in terms of the $v_n$
parameters is perhaps the rapidity dependence of the $v_1$ directed flow; a feature already observed in experiment. However, in itself it is not enough to
determine the tilt angle of the single-particle spectrum.

\section{Summary and outlook}

We have presented a class of rotating and expanding, self-similar solutions of non-relativistic hydrodynamics that describes the expansion of a triaxial 
ellipsoid with non-zero initial angular momentum. Both the initial angular momentum and the triaxial geometry are realistic features, when one considers
the rotating expansion of a hot and dense, strongly interacting matter produced in non-central heavy-ion collisions. The solution presented in this paper 
can accomodate various equations of state, and is a natural generalization of earlier results describing rotating spheroids and non-rotating triaxial
ellipsoids. 

We have evaluated single-particle spectra, flow (azimuthal anisotropy) parameters, two-particle Bose-Einstein correlations for this solution,
using simple formulas that straightforwardly mirror the effect of rotation on the final state observables. The generality in the presented solution (that it
allows for ellipsoids with three different principal axes) makes it possible to draw conclusions on the final state tilt angle, a quantity that in the case
of spheroidal rotating solutions is either ill-defined or at least does not translate into the final state observables. This tilt angle is expected
to behave non-monotonically around the softest point of the equation of state, which may correspond to the critical endpoint of QCD phase transitions.

Although in terms of observables we restricted ourselves to the exact non-relativistic solution discussed in the paper, we are confident that the general
insight our treatment gives into the behavior of them will be useful in analyzing experimental data on single-particle spectra and HBT correlations. A
relativistic generalization, although presently lacking as an actual hydrodynamical solution, maybe derived as a parametrization in the framework of the
Buda-Lund hydrodynamical model. We made some preliminary remarks on this possibility throughout the paper.

A surprising finding of our calculations of observable quantities is that at mid-rapidity, we recover the universal scaling form of the elliptic flow,
$v_2=I_1(w)/I_0(w)$, even for the considered triaxial, rotating and hydrodynamically expanding ellipsoids, as already obtained in Ref.~\cite{Csorgo:2001xm},
for non-relativistic and in Ref.~\cite{Csanad:2005gv} for relativistic kinematic domains. In other words, triaxial, exploding and rotating ellipsoidal
solutions of hydrodynamics do not spoil the universal scaling of the elliptic flow, but they lead to the modification of the definition of the scaling
variable $w$. We also emphasize that the elliptic flow $v_2$ is a dimensionless quantity, hence its universal scaling variable $w$ must also be a
dimensionless, just as is the case in our calculations.

We pointed out that in the general case of rotating \emph{and} tilted source (on which our exact ellipsoidal hydrodynamical solution gives a fairly
reasonable picture) the tilt angle of the ellipsoidal system is not identical to the tilt angle of the single-particle spectrum or that of the
HBT correlation function. We derived expressions that connect these variables, and demonstrated their time dependence (and their values taken at the
freeze-out of the hydrodynamical evolution) for a reasonable set of initial conditions, although a more detailed investigation of this dependence is
beyond the scope of this paper.

In particular, we argued that the tilt angle observable from the oscillating HBT radii and from the single-particle spectrum becomes a non-monotonic
function at the softest point of the equation of state, which suggest that this variable will be useful and straightforward to measure observable
to signal the QCD critical point. We have also found that in the Longitudinal Center of Mass System of the boson pairs the out-long and the side-long
cross-terms oscillate for rotating and expanding triaxial hydrodynamical systems and they are phase-shifted by $\pi/2$ as compared to one another,
providing straighforward experimental testing possibilitiy for the qualitative features obeyed by in our new, rotating and expanding, triaxial
ellipsoidal hydrodynamical solutions.

So the measurement of the rotation by means of the observables discussed in the paper might well give new insights into the equation of state of the
quark-gluon plasma produced in high energy heavy-ion collisions, a main goal in today's heavy ion physics research. We thus look forward to the systematic
experimental exploration of the rotation of the system produced in heavy ion collisions as a function of colliding beam energy, that might reveal a
non-monotonic behavior of the equation of state as a function of temperature and baryochemical potential, and thus might help to locate and
better characterize the deconfinement phase transition.

\section*{Acknowledgements}

We would like to thank to M.~Csan\'ad, L.~P.~Csernai, Y.~Hatta, and K. Ozawa for inspiring discussions. The research of M.~N. has been supported by a
Fulbright Research Grant for the 2015/2016 academic year, as well as by the Hungarian Academy of Sciences through the ``Bolyai J\'anos'' Research
Scholarship program. This work has been supported by the OTKA NK101438 grant of the Hungarian National Science Fund and by a KEK Visitor Fund. 

\appendix
\section{General linear rotating solution}\label{s:app:generalsol}

At the end of Section~\ref{ss:conserved}, we stated that the presented solution follows from the ansatz of a linear velocity field, and the requirement
that it describes self-similarly expanding ellipsoids in the $K'$ frame that is rotating around the $y$ axis. In this Appendix we elucidate this statement,
and write up the most general solution following from this ansatz. After exploring the general solution, we argue that the generality beyond that presented
in the body of the paper is not relevant for the physical problem at hand.

The starting point is the \r{e:roteuler} Euler equation in the $K'$ frame, the \r{e:inertialfdef} expression of the $\v f'$ inertial force density,
the \r{e:sdef} definition of the scaling variable $s$. We require the velocity field to be a generalization of the directional Hubble flow,
to mirror the rotational nature of the flow, to be ``compatible'' (in the sense of \Eq{e:tdst}) with the ellipsoidal scaling variable $s$ of \Eq{e:sdef},
and to be linear in the coordinates in the $K'$ frame. (This implies linearity also in the $K$ frame, but the calculations are easier in the $K'$ frame).
The most general velocity field satisfying these requirements is
\begin{equation}\label{e:vansatz}
\v v'\z{\v r',t} = \z{\begin{array}{l}
\frac{\dot X(t)}{X(t)}r'_x + g(t)\frac{X(t)}{Z(t)}r'_z\vspace{2mm} \\
\frac{\dot Y(t)}{Y(t)}r'_y\vspace{2mm} \\
\frac{\dot Z(t)}{Z(t)}r'_z - g(t)\frac{Z(t)}{X(t)}r'_x \end{array}} ,
\end{equation}
where $X(t)$, $Y(t)$, and $Z(t)$ are the axes of the ellipsoid as in \Eq{e:sdef}, and $g(t)$ is an (up to now) arbitrary function of time.
Now we can readily write up the solutions for the continuity equations (for the number density and the temperature) as in \Eq{e:nTsolA};
with the knowledge of the foregoing examples of similar solutions~\cite{Csorgo:2001ru,Csorgo:2013ksa}, this does not need additional explanation.

The remaining equation to be solved is the Euler equation, \Eq{e:roteuler}. Calculating the $\v f'$ interial force from \Eqs{e:inertialfdef}{e:vansatz}
is straightforward, as is the derivatives of $T$ and $n$; one can plug these into \Eq{e:roteuler}. In order to have a proper solution for the Euler
equation, the coordinate dependence on both sides of it must be identical. This yields that the $\nu(s)$ and $\c T(s)$ functions in \Eq{e:nTsolA}
are not independent, but must obey the condition \r{e:nutau}: if it would not hold, it would be impossible to satisfy the Euler equation
for all spatial coordinates with these velocity, temperature and density fields. But if \r{e:nutau} holds, then the coordinate dependence
of the Euler equation becomes simple: the $x'$ component of the Euler equation will contain terms proportional to $r'_x$ and terms proportional
to $r'_z$, the $y$ component only yields terms proportional to $r'_y$, and the $z'$ components also contain terms proportional to $r'_x$ and $r'_z$.
For all of these to be satisfied for any $\v r'$, we thus get five \emph{ordinary} differential equations for $X(t)$, $Y(t)$, $Z(t)$,
$\dot\vartheta(t)$ and $g(t)$. After some calculation, these turn out to be
\begin{align}
-g^2+\frac{\ddot X}{X}&= \frac{T_0}{m_0}\z{\frac{V_0}{V}}^{\rec\kappa}\rec{X^2} + 2\frac ZXg\dot\vartheta +\dot\vartheta^2,\label{e:app:ddotX}\\
\frac{\ddot Y}{Y}     &= \frac{T_0}{m_0}\z{\frac{V_0}{V}}^{\rec\kappa}\rec{Y^2},\\
-g^2+\frac{\ddot Z}{Z}&= \frac{T_0}{m_0}\z{\frac{V_0}{V}}^{\rec\kappa}\rec{Z^2} + 2\frac XZg\dot\vartheta +\dot\vartheta^2,\label{e:app:ddotZ}
\end{align}
\begin{align}
\frac XZ\dot g+2g\frac{\dot X}{Z} &= -\ddot\vartheta - 2\frac{\dot Z}{Z}\dot\vartheta ,\label{e:app:XZ}\\
\frac ZX\dot g+2g\frac{\dot Z}{X} &= -\ddot\vartheta - 2\frac{\dot X}{X}\dot\vartheta .\label{e:app:ZX}
\end{align}
If $X(t) \neq Z(t)$, then these last two equations can be cast into the form:
\begin{align}
\td{}{t}\sz{\z{X+Z}^2\z{g+\dot\vartheta}} &= 0,\\ 
\td{}{t}\sz{\z{X-Z}^2\z{g-\dot\vartheta}} &= 0,
\end{align}
whose solutions are easily written up as
\begin{align}
g(t)             =& \frac{\chi_0}{\z{X+Z}^2}+\frac{\xi_0}{\z{X-Z}^2},\label{e:app:gsol0}\\
\dot\vartheta(t) =& \frac{\chi_0}{\z{X+Z}^2}-\frac{\xi_0}{\z{X-Z}^2},\label{e:app:dotthetasol0}
\end{align}
with $\chi_0$ and $\xi_0$ constants. Substituting these expressions back into \Eqsdash{e:app:ddotX}{e:app:ddotZ}, we get
\begin{align}
X\ddot X &= \frac{T_0}{m_0}\z{\frac{V_0}{V}}^{\rec\kappa}+ \frac{2\chi_0^2X}{\z{X+Z}^3}+\frac{2\xi_0^2X}{\z{X-Z}^3},\label{e:app:XX}\\
Y\ddot Y &= \frac{T_0}{m_0}\z{\frac{V_0}{V}}^{\rec\kappa},\\
Z\ddot Z &= \frac{T_0}{m_0}\z{\frac{V_0}{V}}^{\rec\kappa}+ \frac{2\chi_0^2Z}{\z{X+Z}^3}-\frac{2\xi_0^2Z}{\z{X-Z}^3}\label{e:app:ZZ}.
\end{align}
Now these equations for $X$, $Y$, $Z$ can be written as the canonical equations from the following Hamiltonian:
\begin{align}
H &= \rec{2m_0}\z{P_X^2+P_Y^2+P_Z^2}+U,\label{e:app:Hamilton}\\
U &= \kappa T_0\z{\frac{V_0}{V}}^{1/\kappa} +\frac{m_0\chi_0^2}{\z{X+Z}^2}+\frac{m_0\xi_0^2}{\z{X-Z}^2} .\label{e:app:Upot}
\end{align}
The most general solution of the hydrodynamical equations with the conditions stated at the beginning of this Appendix is thus given by the formulas
for $\v v'$, $T$ and $n$, with the additional condition that the time evolution of $\dot\vartheta(t)$, $g(t)$ and the axes $X$, $Y$, $Z$ follow
\Eqsdash{e:app:gsol0}{e:app:dotthetasol0} and \Eqsdash{e:app:XX}{e:app:ZZ}. We note that the vorticity of this flow is a slightly more general expression
than in the body of the text, \Eq{e:vorticity}, which is recovered if $\xi_0=0$, i.e. $\dot\vartheta(t) = g(t) \equiv \omega(t)/2$:
\begin{equation}\label{e:app:vorticity}
\gvec\omega_y\z{\v r,t}\equiv \z{\nabla\times\v v\z{\v r,t}}_y = 2\z{\dot\vartheta+g\frac{X^2+Z^2}{2XZ}}.
\end{equation}

We can calculate the conserved quantities for this general solution: the total particle number $N_0$ is again given by \Eq{e:N0}, the total energy
turns out again to be equal to the Hamiltonian \r{e:app:Hamilton}, and the angular momentum $J_z$ now contains a contribution determined by $\xi_0$:
\begin{equation}\label{e:app:Jz}
J_z = N_0m_0\z{\chi_0-\xi_0} .
\end{equation}
The time evolution of a non-rotating ellipsoidal solution is fixed by seven initial conditions: the initial values of the axes and their time derivatives
as well as by the initial temperature $T_0$.
We are now investigating a rotating solution; we expect one additional free initial condition (that corresponds to e.g. the angular momentum of the flow).
The appearance of the \emph{two} new constants, $\chi_0$ and $\xi_0$ in compare to the non-rotating case may thus seem superfluous. This is a justification
to confine ourselves to the case of $\xi_0 = 0$, and in this case the additional initial condition, e.g. the value of $J_z$, is in one-to-one correspondence
with the new constant, $\chi_0$. A more enlightening argument for taking $\xi_0=0$ is that, as seen from the equation of motion of $X$, $Y$, $Z$,
\Eqsdash{e:app:XX}{e:app:ZZ}, in the $\xi_0\neq 0$ case, the potential term \r{e:app:Upot} exhibits an impenetrable potential barrier between the $X>Z$ and
the $X<Z$ regions. So if the initial conditions satisfy $X_0>Z_0$ (as it is physically plausible in the case of a heavy ion collision, see the discussion
in Section~\ref{s:discussion}), then this relation will hold at any future time. On the other hand, realistically one expects that during the time
evolution, because of pressure gradients, we expect that the initially more compressed beam direction, $Z_0 < X_0$ will expand faster 
and eventually in the late stages of the expansion $X<Z$ will hold.

So we conclude that although the $\xi_0\neq 0$ case might be interesting as some exotic rotating expanding flow, it is physically not what we are after
in the quest for the description of a heavy-ion reaction. So we set $\xi_0=0$, and at this point, to get in conformity with the earlier
\cite{Csorgo:2013ksa} result on rotating solutions, we introduce the convenient notation for the $\chi_0$ constant as
\begin{equation}\label{e:app:chi0Romega}
\chi_0 \equiv 2\omega_0R_0^2,\quad R_0\equiv X_0+Z_0.
\end{equation}
With this we get the solution presented in Section~\ref{s:solution}. Of course, for vanishing initial angular momentum we recover the earlier obtained
directional Hubble flow profiles and ellipsoidal exact hydrodynamical solutions.

Returning to the 5 basic equations of motion, \Eqsdash{e:app:ddotX}{e:app:ZX}, it is interesting to see what happens in the $X(t) = Z(t) \equiv R(t)$ case.
In this case \Eqs{e:app:ddotX}{e:app:ddotZ} are the same, so are \Eqs{e:app:XZ}{e:app:ZX}. Four quantities ($R$, $Y$, $\dot\vartheta$, $g$) are
constrained by three equations:
\begin{eqnarray}
\frac{T_0}{m_0}\z{\frac{V_0}{V}}^{1/\kappa} +R^2\z{g+\dot\vartheta}^2
& = & 
R\ddot R 
,\\
\frac{T_0}{m_0}\z{\frac{V_0}{V}}^{1/\kappa} 
& = & 
Y\ddot Y 
, \\
\td{}{t}\sz{R^2\z{g+\dot\vartheta}} & = & 0.
\label{e:app:gthetasum}
\end{eqnarray}
So only the sum, $g+\dot\vartheta$ (which is the total ``angular velocity'' of the fluid) is uniquely determined: in the spheroidal case, one cannot
unequivocally introduce the rotating $K'$ frame and the angular velocity measured in that frame. The remaining freedom in choosing $g$ and
$\dot\vartheta$ can be thought of as some kind of ``gauge freedom''.
\vspace{5mm}
\section{Expression of the new solution in the laboratory frame}
\label{s:app:labsol}
In the body of the paper we have presented our new solution in a frame ($K'$) that rotates together with the ellipsoidal surfaces of the expanding and
rotating fireball. For a concise summary of our solution, let us present ove here the complete solution of the hydrodynamical problem in the laboratory
frame $K$. We strive to write up the formulas in a way that it is easy to compare their forms in the $K$ and $K'$ frames. For clarity, we only present
the solution with homogeneous temperature and Gaussian density profile here (Case B in Section~\ref{ss:hydrosol}). 

First we write up the hydrodynamical equations in Table~\ref{t:app:eqKKprime} in both frames. 
\begin{widetext}
\begin{center}
\begin{table}[h]
\renewcommand{\arraystretch}{1.5}
\begin{tabular}{|c|c|}
\hline
 \textbf{Equations in the laboratory frame} $K$                           & \textbf{Equations in the rotating frame} $K'$ \\\hline
$\partial_t n +\nabla(n\v v)=0$                                           & $\partial'_t n+\nabla'(n \v v') = 0$ \\ 
$\fracd{\m d(\kappa T)}{\m dT}(\partial_t+\v v\nabla)T +T\nabla\v v = 0,$ & $\fracd{\m d(\kappa T)}{\m dT}(\partial'_t+\v v'\nabla')T+T\nabla'\v v'= 0 $ \\
$m_0 n(\partial_t+\v v\nabla)\v v = -\nabla(n T)$                         & $m_0 n (\partial'_t+\v v'\nabla')\v v' = -\nabla'(n T) + \v F'$   \\
   & $\v F' = m_0n\big(2\v v'\times\gvec\Omega+\gvec\Omega\times\z{\v r'\times\gvec\Omega}+\v r'\times\dot{\gvec\Omega}\big)$ \\ \hline
\end{tabular}
\caption{Summary of the hydrodynamical equations for the intertial, laboratory frame $K$ and the same equations in the rotating $K'$ frame, where the
coordinate axes rotate together with a triaxial ellipsoid. The angle of rotation $\vartheta$ and the vector $\gvec\Omega$ are related as $\gvec\Omega =
(0,\dot\vartheta,0)$, as introduced in \Eq{e:Mv}.}\label{t:app:eqKKprime}
\end{table}
\end{center}
\end{widetext}

Now let us summarize the parametric, exact solutions of the hydrodynamical problem presented in the body of the paper, so that the solution is given both
in the laboratory frame $K$ and in the co-rotating frame $K'$. In Section~\ref{ss:hydrosol} the solution was presented in the $K'$ frame, the frame that
fits naturally to the rotating nature of our solution. Using \Eqsdash{e:rxtransform}{e:vztransform}, it is easy to write up the solution in the $K$ frame.
The resulting formulas can be found in Table~\ref{t:app:solK}.
\begin{widetext}
\begin{center}
\begin{table}[h]
\renewcommand{\arraystretch}{1.8}
\begin{tabular}{|cc|c|}
\hline
\multicolumn{3}{|l|}{\bf In both frames:}  \\\hline
$\begin{array}{l}
H_x=\fracd{\dot X}{X}, \quad H_y=\fracd{\dot Y}{Y}, \quad H_z=\fracd{\dot Z}{Z}, \\
V = (2 \pi)^{3/2} XYZ,\quad n = n_0\fracd{V_0}{V}\exp\z{-s/2},
\end{array}$ & \multicolumn{2}{c|}{
  $\begin{array}{lcl}
  \fracd{\m d\sz{T\kappa(T)}}{\m dT}\fracd{\dot T}{T} + \fracd{\dot V}{V}=0 &
  \,\m{if}\, &\kappa(T)\neq\mbox{\it const},\\
  T=T_0\z{\fracd{V_0}{V}}^{1/\kappa}&
  \,\m{if}\, &\kappa(T)=\mbox{\it const} \end{array} $}\\ 
 $\dot\vartheta\equiv\fracd{\omega}{2},\qquad\omega=\omega_0\fracd{R_0^2}{R^2}, \qquad R = \fracd{X+Z}{2}, $ &
\multicolumn{2}{c|}{ $X\big(\ddot X-\omega^2R\big) = Y\ddot Y = Z\big(\ddot Z -\omega^2R\big) = \fracd{T}{m_0},$}\\\hline
\multicolumn{2}{|l|}{\textbf{in laboratory frame} $K$:} & \textbf{in the co-rotating frame} $ K'$:\\\hline 
\multicolumn{2}{|c|}{
$s = \fracd{r_x^2}{X^2}+\fracd{r_y^2}{Y^2}+\fracd{r_z^2}{Z^2} + \z{\recd{Z^2}-\recd{X^2}}\sz{(r_x^2-r_z^2)\sin^2\vartheta+r_xr_z\sin(2\vartheta)} $ }&
$s = \fracd{{r_x'}^2}{X^2}+\fracd{{r_y'}^2}{Y^2}+\fracd{{r_z'}^2}{Z^2}$ \\
\multicolumn{2}{|c|}{
$ \v v\z{\v r,t} = \v v_H\z{\v r,t} + \v v_R\z{\v r,t} $}  &   $ \v v'\z{\v r',t} = \v v'_H\z{\v r',t} + \v v'_R\z{\v r',t} $  \\
\multicolumn{2}{|c|}{
$ \v v_H\z{\v r,t} = \begin{pmatrix} (H_x\m{cos}^2\vartheta+H_z\m{sin}^2\vartheta)r_x \\ H_yr_y \\ (H_x\m{sin}^2\vartheta+H_z\m{cos}^2\vartheta)r_z
                     \end{pmatrix} + (H_z-H_x)\fracd{\sin(2\vartheta)}{2}\begin{pmatrix} r_z\\0\\r_x \end{pmatrix} $  }&
$ \v v'_H\z{\v r',t} =  \begin{pmatrix} H_xr'_x \\ H_yr'_y \\ H_zr'_z \end{pmatrix}$ \\
\multicolumn{2}{|c|}{
$ \v v_R\z{\v r,t} = 
                      \dot\vartheta \begin{pmatrix} r_z \\0\\ - r_x \end{pmatrix}
                     + \dot\vartheta \begin{pmatrix} \z{\fracd{X}{Z}\m{cos}^2\vartheta+ \fracd{Z}{X} \m{sin}^2\vartheta} r_z \\0\\ -\z{\fracd{X}{Z}\m{sin}^2\vartheta
                                    +\fracd{Z}{X} \m{cos}^2\vartheta} r_x \end{pmatrix}
                     + \dot\vartheta \left(\fracd{X}{Z} - \fracd{Z}{X}\right) \fracd{\sin(2\vartheta)}{2}\begin{pmatrix} r_x\\0\\-r_z \end{pmatrix} $
  } &
$ \v v'_R\z{\v r',t} = \dot\vartheta\begin{pmatrix} \fracd{ X}{Z} r'_z \\0\\ -\fracd{Z}{X} r'_x \end{pmatrix} $ \\  \hline
\end{tabular}
\caption{
Summary of the new rotating solution of the hydrodynamical equations, written up both in the intertial, laboratory frame $K$ and in the $K'$ frame,
where the coordinate axes rotate together with the $(X,Z)$ axes of a triaxial ellipsoid.}
\label{t:app:solK}
\end{table}
\end{center}
\end{widetext}
The dynamical equations that describe the time evolution of the scale parameters $(X,Y,Z)$ and the temperature $T$ are the same both in $K$ and in $K'$.

We have written up the velocity field as a sum of two terms: a ,,Hubble-term'' $\v v_H$, and a ,,rotational term'' $\v v_R$. The directional Hubble flow
and its Hubble constants $(H_x, H_y, H_z)$ have a very clear meaning in the rotating frame $K'$, where the Hubble component of the velocity field, $v_H$
is diagonal; this is not the case in the $K$ frame. The distinction between $\v v_H$ and $\v v_R$ is that the Hubble term has zero curl (and thus does
not contribute to the vorticity of the flow), while the rotational term has zero divergence. So for the divergence we can write: 
\begin{equation}
\nabla \v v = \nabla \v v_H = \nabla' \v v'_H = \frac{\dot V}{V},\quad \nabla \v v_R = \nabla'\v v'_R = 0. 
\end{equation}
The terms in $\v v$ that are proportional to the angular velocity $\dot\vartheta$ contribute to the rotational flow velocity $\v v_R$, which determines
the vorticity of the solution as
\begin{align}
\gvec\omega\z{\v r,t}   &\equiv \nabla \times\v v  = \nabla \times\v v_R  ,\quad &\nabla\times \v v_H = 0,\\
\gvec\omega'\z{\v r',t} &\equiv \nabla'\times\v v' = \nabla'\times\v v'_R ,\quad &\nabla\times\v v'_H = 0.
\end{align}
The vorticity vector is parallel with the axis of rotation, and the value of its only non-vanishing component in the laboratory frame, 
$\gvec\omega_y$, was given already in \Eq{e:vorticity}. We can also write up it in the rotating $K'$ frame; we have
\begin{align}
\gvec \omega_y\z{\v r,t}   &= \omega + \frac \omega 2\z{\frac XZ +\frac ZX},\\
\gvec\omega_y'\z{\v r', t} &= \frac \omega 2\z{\frac XZ +\frac ZX}.
\end{align}
In the $X\rightarrow Z$ limit, it is easy to confirm from Table~\ref{t:app:solK} that $\omega=2\dot\vartheta$ is indeed the angular velocity of the fluid.

Thus Table ~\ref{t:app:solK} summarizes our new solutions for the case of the spatially homogeneous temperature profile. This class of
solutions allows for a temperature dependent (but otherwise general, unrestricted) $\kappa(T)$ function. Again, we note that a solution with arbitrary
temperature profile and corresponding density profile was also given in Section~\ref{ss:hydrosol}, but for simplicity they are not included in
Table~\ref{t:app:solK} of this Appendix.

\end{document}